\documentclass[aip,jmp,preprint]{revtex4-1}

\usepackage{amsfonts,amssymb,amsmath,graphicx}
\usepackage{verbatim,xcolor}
\date{\today}

\def\iu{{\rm i}}

\def\bfe{\boldsymbol E}
\def\bfb{\boldsymbol B}

\def\bas{{\boldsymbol A}^{\text{sc}}}
\def\bjs{\boldsymbol{\mathfrak J}}
\def\bsm{\boldsymbol{\mathcal S}}
\def\bkm{\boldsymbol{\Lambda}}

\def\ez{\mathbf e_z}
\def\ex{\mathbf e_x}
\def\ey{\mathbf e_y}

\def\dxp{{\rm d}x'}
\def\dx{{\rm d}x}

\def\dxi{{\rm d}\xi}
\def\dxip{{\rm d}\xi'}

\begin{document}

\title{Edge plasmon-polaritons on isotropic semi-infinite conducting sheets}

\author{Dionisios Margetis}
\email{dio@math.umd.edu}

\affiliation{Department of Mathematics, and Institute for Physical Science
and Technology, and Center for Scientific Computation and Mathematical Modeling, University of Maryland, College Park, Maryland 20742, USA}

\begin{abstract}
From a three-dimensional boundary value problem for the time harmonic classical Maxwell equations, we derive the dispersion relation for a surface wave, the \emph{edge plasmon-polariton} (EP), that is localized near and propagates along the straight edge of a planar, semi-infinite sheet with a {\em spatially homogeneous, scalar conductivity}. The sheet lies in a uniform and isotropic medium; and serves as a model for some two-dimensional (2D) conducting materials such as the doped monolayer graphene. We formulate a homogeneous system of integral equations for the electric field tangential to the plane of the sheet. By the Wiener-Hopf method, we convert this system to coupled functional equations on the real line for the Fourier transforms of the fields in the surface coordinate normal to the edge, and solve these equations exactly. The derived EP dispersion relation smoothly connects two regimes: a low-frequency regime, where the EP wave number, $q$, can be comparable to the propagation constant, $k_0$, of the ambient medium; and the nonretarded frequency regime in which $|q|\gg |k_0|$. Our analysis indicates two types of 2D surface plasmon-polaritons on the sheet away from the edge. We extend the formalism to the geometry of two coplanar sheets.
\end{abstract}




\maketitle

\section{Introduction}
\label{sec:Intro}
Research efforts in the design and fabrication of two-dimensional (2D) materials rapidly evolved into a rich field at the crossroads of physics, chemistry, materials science and engineering.~\cite{Torres2014,Geimetal2013} Some of these materials, including graphene and black phosphorus, are highly promising ingredients of  nanophotonics at the mid- and near-infrared frequencies.~\cite{Zhang2010} These systems may possibly sustain evanescent, fine-scale electromagnetic waves that are tightly confined to the boundary.~\cite{Lowetal2017,Pitarkeetal2007, Gonzalez2014,Luetal2016} An appealing surface wave is the 2D ``bulk'' surface plasmon-polariton (SP), which expresses collective excitations of the electron charge in the 2D plasma.~\cite{Hanson2008,Bludov2013,Jablan2013} The SP may exhibit wavelengths much shorter than those in the ambient dielectric medium, and thus may overcome the typical diffraction limit.\cite{Zhang2012,Huidobro2010} 

Experimental observations suggest that 2D bulk SPs on graphene nanoribbons are accompanied by different short-scale waves, termed ``edge plasmon-polaritons'' (EPs), which oscillate rapidly along the edges of the 2D material.~\cite{Feietal2015,Yanetal2012,Crasseeetal2012,Taoetal2011} The EP is localized near each edge on the sheet and may have a wavelength shorter than the one of the accompanying bulk SP at terahertz frequencies. This EP is intimately related to the edge magnetoplasmon which was observed to propagate along the boundary of the electron layer on liquid ${}^4$He in the presence of a static magnetic field.~\cite{Fetter1985,MastFetter1985,VolkovMikhailov1985} Aspects of this wave have been studied via linear models for confined 2D electron systems.~\cite{Fetter1986,Fetter1986b,Wuetal1985,VolkovMikhailov1988,CohenGoldstein2018} To our knowledge, many studies of the EP are restricted to the nonretarded frequency regime, in which the electric field is approximated by the gradient of a scalar potential (``quasi-electrostatic approach'').\cite{Pitarkeetal2007,MMSLL-preprint}

In this paper, we use the time harmonic classical Maxwell equations in three spatial dimensions (3D) in order to formally derive the dispersion relation of the EP on a semi-infinite, flat sheet with a straight edge and a homogeneous and isotropic surface conductivity. The sheet lies in a homogeneous and isotropic medium. We formulate a system of integral equations for the electric field tangential to 
the plane of the sheet; and apply the Wiener-Hopf method to solve these equations exactly via the Fourier transform in the sheet coordinate normal to the edge. In this sense, our treatment accounts for retardation effects. The EP dispersion relation is derived through the analyticity of the Fourier-transformed fields. 

Our tasks and results are summarized as follows.

\begin{itemize}

\item We formulate a boundary value problem for time harmonic Maxwell's equations in 3D. An ingredient is a jump condition for the 
tangential magnetic field across the sheet with an assumed local and tensor-valued surface conductivity.\cite{Bludov2013,Hanson2008}

\item We convert this boundary value problem to a system of coupled integral equations for the electric field components tangential to the plane of the sheet. The kernel comes from the retarded Green function for the vector potential. 

\item We apply the Wiener-Hopf 
method~\cite{Krein1962,WienerHopf1931,PaleyWiener,Masujima-book} when the sheet is spatially homogeneous.  In this context, we use the Fourier transform of the fields in the surface coordinate normal to the edge, and convert the integral equations to functional equations on the real line. 

\item For isotropic sheets, we formally solve these functional equations exactly via a suitable linear transformation of the tangential electric field. 

\item We derive the EP dispersion relation via enforcing the requisite analyticity of the Fourier-transformed fields. The ensuing relation exhibits the joint contributions of transverse-magnetic (TM) and transverse-electric (TE) field polarizations.

\item
For a given EP wave number, we describe 2D SPs in the direction vertical to the edge. 

\item For a fixed phase of the surface conductivity with recourse to the Drude model,\cite{Jablan2013} we derive an asymptotic expansion for the EP wave number at low enough frequencies. This expansion provides a refined description of the gapless EP energy spectrum.\cite{VolkovMikhailov1988}

\item We compare the derived EP dispersion relation to the respective result of the quasi-electrostatic approach.~\cite{VolkovMikhailov1988} We extract the leading-order correction due to retardation.

\item We provide an extension of our analysis to the geometry with two coplanar, semi-infinite sheets of distinct isotropic, spatially homogeneous surface conductivities.

\end{itemize}

We should mention the models of hydrodynamic flavor for magnetoplasmons found in Refs.~\onlinecite{MastFetter1985,Fetter1985,Wuetal1985,Fetter1986,Fetter1986b}. The main idea in these works is to couple the 2D (non-relativistic) linearized Euler equation with the 3D Poisson equation for an electrostatic potential. A dispersion relation for the edge magnetoplasmon is then obtained via an ad hoc simplification of the integral relation between the potential and the electron density; the exact kernel is replaced by a simpler one having the same infrared 
behavior.~\cite{MastFetter1985,Wuetal1985,Fetter1985,Fetter1986,Fetter1986b} This treatment offers insights into the effect of the geometry and the relative importance of bulk and edge contributions; see, e.g., Refs.~\onlinecite{Eliassonetal1986,Cataudella1987,Mikhailov1995,Wangetal2012,Zabolotnykh2016}. A few limitations of these works, on the other hand, are evident. 
For instance, the time harmonic electric field is approximately expressed as the gradient of a scalar potential, which poses a restriction on the magnitude of the EP wave number.  Moreover, the simplified relation between potential and electron density may become questionable, e.g., for the calculation of the EP phase velocity at long wavelengths.~\cite{CohenGoldstein2018} 

Other works of similar, hydrodynamic character invoke the nonlocal mapping from the electron density to the potential along the 2D material, and resort to numerics in the context of the quasi-electrostatic approximation.~\cite{Gumbs1989,Rudin1997,Nikitin2011} Note that in Ref.~\onlinecite{Vaman2014} the integral equations for the oscillation amplitudes of electrons are apparently solved explicitly, without any kernel approximation, yet by the neglect of retardation; see also Ref.~\onlinecite{ApostolVaman2009}. An extension of these treatments to viscous electron flows via the Wiener-Hopf method is found in Ref.~\onlinecite{CohenGoldstein2018}.

The dispersion relation of edge magnetoplasmons has been systematically derived from an anti-symmetric tensor surface conductivity via the solution of an integral equation for the electrostatic potential by the Wiener-Hopf method.~\cite{VolkovMikhailov1985,VolkovMikhailov1988} In these works, the quasi-electrostatic approximation  is applied from the start. In contrast to Refs.~\onlinecite{MastFetter1985,Wuetal1985,Fetter1985,Fetter1986,Fetter1986b,VolkovMikhailov1985,VolkovMikhailov1988,Eliassonetal1986,Cataudella1987,Mikhailov1995,Wangetal2012,Zabolotnykh2016,CohenGoldstein2018,Gumbs1989,Rudin1997,Nikitin2011,Vaman2014}, we solve the full Maxwell equations here, albeit in isotropic settings. We address the cases with a single sheet and two coplanar sheets with homogeneous scalar conductivities. 

Our approach is motivated by the need to describe the dispersion of plasmon-polaritons in a wide range of frequencies and 2D materials.~\cite{Lowetal2017} We obtain the EP dispersion relation in the form $\mathcal F(q,\omega)=0$, where $q$ is the EP wave number, $\omega$ is the frequency and $\mathcal F$ is a transcendental function that, for a given surface conductivity function $\sigma(\omega)$, smoothly connects: the nonretarded frequency regime, in which $q/\omega^2\simeq {\rm const.}$;\cite{Bludov2013} and the low-frequency regime, in which $q/\omega\simeq {\rm const.}$. We provide corrections to these leading-order terms by assuming that $\Im\,\sigma(\omega)>0$; this condition is consistent with the Drude model for $\sigma(\omega)$.\cite{Jablan2013}

In this vein, we analytically show how the EP dispersion relation bears the signatures of both the TM and TE polarizations. In particular, the contribution of the TE-polarization becomes relatively small in the quasi-electrostatic limit, for $\Im\,\sigma(\omega)>0$.

Our work points to several open questions. Our approach, relying on the solution of the full Maxwell equations, does not address anisotropic and nonlocal effects in the surface conductivity.~\cite{Pitarkeetal2007,CohenGoldstein2018} Tensor-valued, spatially constant surface conductivities in principle can lead to challenging systems of Wiener-Hopf integral equations for the electric field.~\cite{GohbergKrein1960,WuWu1963,Abrahams1997} We also neglect the effect that the edge, as a boundary of a 2D electron system, has on the conductivity. The geometry of the semi-infinite conducting sheet is not too realistic. The experimentally appealing case of nanoribbons will be the subject of future work. Since we focus on analytical aspects of EPs, numerical predictions will be addressed elsewhere.\cite{MMSLL-inprep}  

\subsection{Outline}
\label{subsec:outline}

The remainder of this paper is organized as follows. In Section~\ref{sec:summary}, we summarize our key results for isotropic sheets. \color{black} In Section~\ref{sec:formulation}, we state the boundary value problem (Section~\ref{subsec:geom-bvp}); and formulate integral equations for the electric field on a flat sheet in a homogeneous isotropic medium (Section~\ref{subsec:integral}). Section~\ref{sec:W-H} describes the coupled functional equations for the Fourier transforms of the electric field components tangential to a homogeneous sheet. In Section~\ref{sec:solution}, we use a homogeneous scalar conductivity to: obtain decoupled functional equations via a linear field  transformation (Section~\ref{subsec:explicit}); and derive the EP dispersion relation (Section~\ref{subsec:EP-disp}). In Section~\ref{subsec:EP_and_SP} we compute the tangential electric field, and describe the 2D bulk SPs in the direction normal to the edge. In Section~\ref{sec:small-q}, we simplify the EP dispersion at low frequencies. Section~\ref{sec:qs} focuses on the asymptotics related to the quasi-electrostatic approximation. In Section~\ref{sec:extension}, we extend our analysis to two coplanar isotropic sheets. Section~\ref{sec:conclusion} concludes the paper with a discussion of open problems. 

\subsection{Notation and terminology}
\label{subsec:notation}

In our analysis, $\mathbb{C}$ is the complex plane, $\mathbb{R}$ is the set of real numbers and $\mathbb{Z}$ is the set of integers. $w^*$ is the complex conjugate of $w$ ($w\in \mathbb{C}$). $\Re w$ ($\Im w$) denotes the real (imaginary) part of complex $w$. Boldface symbols denote vectors or matrices; e.g., $\mathbf e_\ell$ is the $\ell$-directed unit Cartesian vector $(\ell= x, y, z)$. The Hermitian part of matrix $\boldsymbol M$ is $\frac{1}{2}({{\boldsymbol M}^*}^T+\boldsymbol M)$ where the asterisk $({}^*)$ and $T$ as superscripts denote complex conjugation and transposition, respectively. \color{black} We write $f=\mathcal O(g)$ ($f=o(g)$) to mean that $|f/g|$ is bounded by a nonzero constant (approaches zero) in a prescribed limit; and  $f\sim g$ implies $f-g=o(g)$. The term ``sheet'' means  {\em either} a material thin film, {\em or} a Riemann sheet as a branch of a multiple-value function. The terms ``top Riemann sheet'' and ``first Riemann sheet'' are employed interchangeably; ditto for the terms ``wave number'' and ``propagation constant''. Given a function, $F(\xi)$, of a complex variable, $\xi$, we define the functions $F_\pm(\xi)$ by $F(\xi)=F_+(\xi)+F_-(\xi)$ where (i) $F_+(\xi)$ is analytic in the upper half $\xi$-plane, $\mathbb{C}_+=\{\xi\in\mathbb{C}: \Im\,\xi>0\}$; and (ii) $F_-(\xi)$ is analytic in the lower half $\xi$-plane, $\mathbb{C}_-=\{\xi\in\mathbb{C}: \Im\,\xi<0\}$. The $e^{-\iu\omega t}$ time dependence is used throughout where $\omega$ is the angular frequency, and $\omega>0$ unless we state otherwise ($\iu^2=-1$). We employ the International System of units (SI units) throughout.

\section{Main results}
\label{sec:summary}

In this section, we summarize our key results regarding isotropic sheets. The derivations can be found in corresponding sections as specified below.

Suppose that the conducting material is the set $\Sigma=\{(x,y,z)\in\mathbb{R}^3: x>0, z=0\}$, and has scalar surface conductivity $\sigma(\omega)$ in the frequency domain. Thus, the material edge is identified with the $y$-axis. The sheet is surrounded by an isotropic and homogeneous medium. To study the EP dispersion, suppose that all fields have the $e^{\iu qy}$ dependence on the $y$ coordinate, where the complex $q$ needs to be determined as a function of $\omega$.

\subsection{Integral equations for EP electric field tangential to isotropic sheet}
\label{subsec:summ-IE-edge}
The electric field parallel to the plane of the sheet is of the form $e^{\iu q y}\boldsymbol E_\parallel(x,z)$ where $\boldsymbol E_\parallel(x,z)=(\widetilde E_x(x,z), \widetilde E_y(x,z),0)^T$. This $\boldsymbol E_\parallel(x,z)$ is continuous across the sheet. We show that, in the absence of any incident field, $\boldsymbol E_\parallel(x,z)$ at $z=0$ satisfies 
\begin{equation}\label{eq:integral-reln-Epar}
\boldsymbol E_\parallel(x,0)
=    \frac{\iu\omega\mu\sigma}{k_0^2}
\begin{pmatrix}
\displaystyle \frac{{\rm d}^2}{\dx^2}+k_0^2 &  \quad \displaystyle \iu q \frac{{\rm d}}{\dx} \\
  \displaystyle \iu q\frac{{\rm d}}{\dx}      &   k_{\rm eff}^2   	
\end{pmatrix} 
\int_0^\infty \dxp\, K(x-x';q)\,\boldsymbol E_\parallel(x',0)~, \quad  \mbox{all}\ x\ \mbox{in}\ \mathbb{R}~.          	
\end{equation}
Here, we ignore the (zero) $z$-component of $\boldsymbol E_\parallel$, and define $k_0=\omega\sqrt{\varepsilon\mu}$ where $\varepsilon$ and $\mu$ are the dielectric permittivity and magnetic permeability of the ambient medium, respectively; and $k_{\rm eff}^2=k_0^2-q^2$ with $\Im\,k_{\rm eff}>0$. The kernel is $K(x;q)=G(x,0;0,0)$ where $G(x,z;x',z')$ is the retarded Green function for the scalar Helmholtz equation with wave number $k_{\rm eff}$. {\em The EP dispersion relation is saught by requiring that~\eqref{eq:integral-reln-Epar} admit nontrivial integrable solutions.}

Equation~\eqref{eq:integral-reln-Epar} is a particular case of the integral system obtained when the surface conductivity is anisotropic; see Section~\ref{subsec:integral}. The derivation of~\eqref{eq:integral-reln-Epar} and its extension is described in Section~\ref{subsec:integral}. In Section~\ref{sec:W-H}, we use the Fourier transform in $x$ in order to state the respective matrix Riemann-Hilbert problem.

\subsection{EP dispersion relation for isotropic sheet}
\label{subsec:summ-EP-disp-1sheet}

Without loss of generality, assume that $\Re\,q>0$. By~\eqref{eq:integral-reln-Epar} the EP dispersion relation is
\begin{subequations}
\begin{equation}\label{eq:analt-EP-disp}
\exp\left\{
\left[Q_+(\iu q)+Q_-(-\iu q)\right]-\left[R_+(\iu q)+R_-(-\iu q)\right]\right\}=-1~;
\end{equation}
for $\Re\,q<0$ simply replace $q$ by $-q$ in this relation. In the above, we define
\begin{equation}\label{eqs:summ-QR-def}
Q_\pm(\xi)=\pm\frac{1}{2\pi\iu}\int_{-\infty}^\infty \frac{\ln\mathcal P_{\rm TM}(\xi')}{\xi'-\xi}\,{\rm d}\xi'~,\quad 
R_\pm(\xi)=\pm\frac{1}{2\pi\iu}\int_{-\infty}^\infty \frac{\ln\mathcal P_{\rm TE}(\xi')}{\xi'-\xi}\,{\rm d}\xi'~,\ \pm\Im\,\xi>0~;
\end{equation}	
\begin{equation}\label{eqs:summ-P-def}
\mathcal P_{\rm TM}(\xi)=1-\frac{\iu \omega\mu\sigma}{k_0^2}(k_{\rm eff}^2-\xi^2)\widehat K(\xi;q)~,\quad 
\mathcal P_{\rm TE}(\xi;q)=1-\iu\omega\mu\sigma\widehat K(\xi;q)~.
\end{equation}
\end{subequations}
Here, $\widehat K(\xi;q)$ is the Fourier transform of kernel $K(x;q)$; $\widehat K(\xi;q)=(\iu/2)(k_{\rm eff}^2-\xi^2)^{-1/2}$ with $\Im\sqrt{k_{\rm eff}^2-\xi^2}>0$ for wave decay in $|z|$. The derivation of~\eqref{eq:analt-EP-disp} is provided in Section~\ref{sec:solution}.

\subsection{Electric field tangential to plane of sheet near and away from edge}
\label{subsec:near-far-E}
Suppose that $q$ satisfies~\eqref{eq:analt-EP-disp}. By using the ensuing Fourier integrals for the components of $\boldsymbol E_\parallel(x,0)$, we show that $\mathbf{e}_x\cdot \boldsymbol E_\parallel(x,0)$ is singular at the edge, viz.,
\begin{equation*}
	\mathbf{e}_x\cdot \boldsymbol E_\parallel(x,0)=\mathcal O(\sqrt{k_0 x})\ \mbox{as}\ x\downarrow 0\quad \mbox{and}\quad \mathbf{e}_x\cdot \boldsymbol E_\parallel(x,0)=\mathcal O((k_0 x)^{-1/2})\ \mbox{as}\ x\uparrow 0~;
\end{equation*}
whereas $\mathbf{e}_y\cdot \boldsymbol E_\parallel(x,0)$ is continuous and finite at the edge.
For the derivations see Section~\ref{subsec:near-f}.

For the far field on the sheet (as $x\to +\infty$ for $z=0$) we write $\boldsymbol E_\parallel=\boldsymbol E_\parallel^{\rm sp}+\boldsymbol E_\parallel^{\rm rad}$; $\boldsymbol E_\parallel^{\rm sp}$ amounts to a 2D bulk SP as the residue contribution to the Fourier integrals from a zero of $\mathcal P_{\rm TM}(\xi)$ or $\mathcal P_{\rm TE}(\xi)$, whereas $\boldsymbol E_\parallel^{\rm rad}$ is the branch cut contribution. We derive asymptotic formulas for  $\boldsymbol E_\parallel^{\rm rad}(x,0)$ and exact formulas for $\boldsymbol E_\parallel^{\rm sp}(x,0)$ for $x>0$.
For example, we find that
\begin{equation*}
	|\boldsymbol E_\parallel^{\rm rad}(x,0)|=\mathcal O\biggl(\frac{e^{-\sqrt{q^2-k_0^2}\,x}}{(\sqrt{q^2-k_0^2} x)^{3/2}}\biggr)\quad \mbox{as}\ |\sqrt{q^2-k_0^2}\,x|\to +\infty~,
\end{equation*}
keeping $q/k_0$ and $\omega\mu\sigma/k_0$ fixed. For details, see Section~\ref{subsec:far-f}. In a similar vein, we have
\begin{equation*}
	|\boldsymbol E_\parallel^{\rm sp}(x,0)|=\mathcal O\bigl(e^{\iu k_{\rm sp}x}\bigr)
\end{equation*}
where, for a lossless ambient medium ($k_0>0$), $k_{\rm sp}$ is the zero in the upper half $\xi$-plane 
of $\mathcal P_{\rm TM}(\xi)$ if $\Im\,\sigma>0$, or $\mathcal P_{\rm TE}(\xi)$ if $\Im\,\sigma<0$; see Section~\ref{subsec:far-f}.

\subsection{Approximation for EP dispersion relation at low frequency}
\label{subsec:summ-low-freq}

If $|\omega\mu\sigma(\omega)/k_0|\gg 1$ along with $\Im\,\sigma(\omega)>0$, we show that~\eqref{eq:analt-EP-disp} yields the approximation
 \begin{equation*}
 	\frac{q-k_0}{k_0}\sim \frac{\epsilon^2}{2\pi^2}\mathcal A(\epsilon)^2\quad \mbox{where}\quad e^{\mathcal A(\epsilon)}=\frac{2e\pi}{\epsilon^2\mathcal A(\epsilon)}~;\quad \epsilon=\frac{\iu\,2k_0}{\omega\mu\sigma}\quad (|\epsilon|\ll 1)~.
 \end{equation*}
For details, see Section~\ref{sec:small-q}. In view of the semi-classical Drude model\cite{Jablan2013} for $\sigma(\omega)$, the above asymptotic formula indicates how $q$ approaches $k_0$ at low enough frequency, $\omega$.

\subsection{EP dispersion relation in nonretarded regime}
\label{subsec:summ-nonret}
In the nonretarded frequency regime, when $|\omega\mu\sigma(\omega)/k_0|\ll 1$ with $\Im\,\sigma(\omega)>0$,\cite{Bludov2013,Jablan2013} the EP dispersion relation can be derived by the quasi-electrostatic approach.\cite{VolkovMikhailov1985,VolkovMikhailov1988,MMSLL-preprint} By carrying out an asymptotic expansion for exact result~\eqref{eq:analt-EP-disp}, we derive the formula
\begin{equation*}
 q\sim \iu \eta_0\,\frac{2k_0^2}{\omega\mu\sigma}\biggl\{1-\eta_1\biggl(\frac{\omega\mu\sigma}{2k_0}\biggr)^2\biggr\}~.
 \end{equation*}
In the above, $\eta_0$ is a numerical factor ($\eta_0\simeq 1.217$) that amounts to the result of the quasi-electrostatic approximation;\cite{VolkovMikhailov1988} and $\eta_1$ is a positive constant ($\eta_1\simeq 0.416$) that signifies the leading-order correction due to retardation; see Section~\ref{sec:qs}.

\subsection{Extension of EP theory to two coplanar conducting sheets}
\label{subsec:summ-2sheets}
Consider the coplanar sheets described by the sets $\Sigma^L=\{(x,y,z)\in\mathbb{R}^3\,:\,z=0,\, x<0\}$ and $\Sigma^R=\{(x,y,z)\in\mathbb{R}^3\,:\,z=0,\, x>0\}$, which lie in an isotropic homogeneous medium. Suppose  that their scalar, spatially constant conductivities are $\sigma^L$ and $\sigma^R$, respectively ($\sigma^{L}\neq \sigma^{R}$ and $\sigma^L\sigma^R\neq 0$). The electric field tangential to the sheets in $\Sigma=\Sigma^L\cup \Sigma^R$ satisfies 
\begin{equation*}
\boldsymbol E_\parallel(x,0) =    \frac{\iu\omega\mu}{k_0^2}
\begin{pmatrix}
\displaystyle \frac{{\rm d}^2}{\dx^2}+k_0^2 &  \quad \displaystyle \iu q \frac{{\rm d}}{\dx} \\
  \displaystyle \iu q\frac{{\rm d}}{\dx}      &   k_{\rm eff}^2   	
\end{pmatrix} 
\int_{-\infty}^\infty \dxp\, K(x-x';q)\sigma(x')\,\boldsymbol E_\parallel(x',0) \quad  \mbox{all}\ x\ \mbox{in}\ \mathbb{R}~,            
\end{equation*}
where $\sigma(x)=\sigma^L+\vartheta(x) (\sigma^R-\sigma^L)$; $\vartheta(x)=1$ if $x>0$ and $\vartheta(x)=0$ if $x<0$ (see Section~\ref{sec:extension}).

 We show that the equation for $\boldsymbol E_\parallel(x,0)$ admits nontrivial integrable solutions if $q$ obeys~\eqref{eq:analt-EP-disp} with definitions~\eqref{eqs:summ-QR-def} for $Q_\pm(\xi)$ and $R_{\pm}(\xi)$. However, in the latter formulas one should make the replacement $\mathcal P_{\rm \varpi}(\xi)\rightarrow \mathcal P_{\varpi}^R(\xi)/\mathcal P_{\varpi}^L(\xi)$  where $\mathcal P_{\varpi}^{\ell}(\xi)$ is defined by~\eqref{eqs:summ-P-def} with $\sigma\rightarrow \sigma^{\ell}$ ($\varpi={\rm TM}, {\rm TE}$ and $\ell=R, L$). See Section~\ref{sec:extension} for some details.

\color{black}

\section{Boundary value problem and integral equations}
\label{sec:formulation}

In this section, we formulate a boundary value problem for time harmonic Maxwell's equations in the geometry with a semi-infinite conducting sheet in an unbounded, isotropic and homogeneous medium. We also derive a system of integral equations for the electric field tangential to the plane of the sheet. Our formulation includes nonhomogeneous and anisotropic sheets with {\em local} surface conductivities; a generalization is provided in Ref.~\onlinecite{MMSLL-preprint}.

\subsection{Geometry and boundary value problem}
\label{subsec:geom-bvp}

\begin{figure}
\includegraphics*[scale=0.4, trim=0in 0.7in 0in 1.5in]{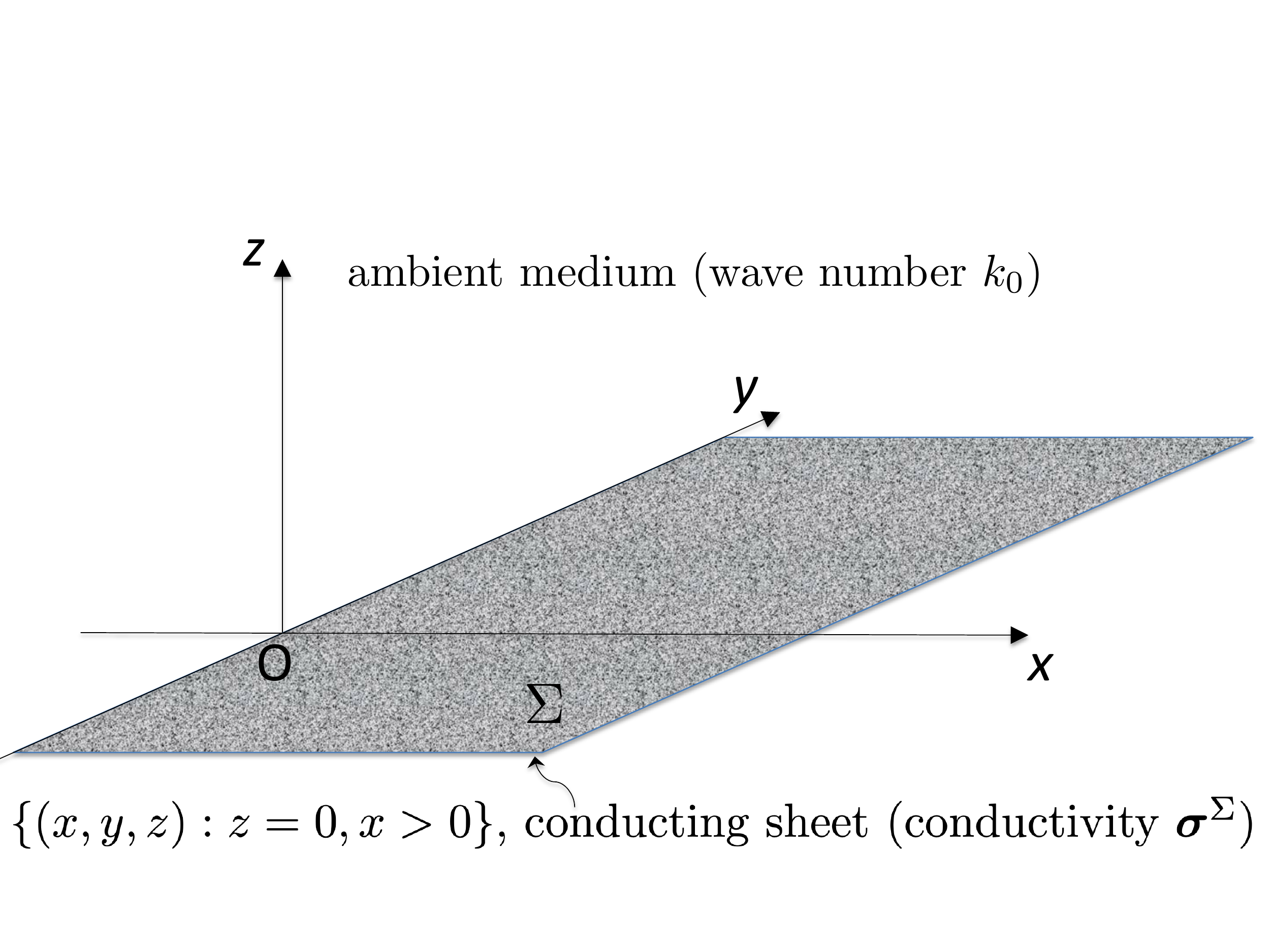}
    \caption{Geometry of the problem. A semi-infinite conducting sheet, $\Sigma$, lies in the $xy$-plane for $x>0$, and has the local and in principle tensor-valued surface conductivity $\boldsymbol\sigma^\Sigma$. The sheet is surrounded by a homogeneous and isotropic medium of wave number $k_0=\omega\sqrt{\varepsilon\mu}$, where $\varepsilon$ is the dielectric permittivity and $\mu$ is the magnetic permeability.}\label{fig:Geom}
\end{figure}

The geometry of the problem is depicted in Fig.~\ref{fig:Geom}. This consists of: a semi-infinite conducting sheet, $\Sigma=\{(x,y,z)\in\mathbb{R}^3: x> 0~,\ z=0\}$, in the $xy$-plane; and the surrounding unbounded, homogeneous and isotropic medium. The sheet, $\Sigma$, has a local and in principle tensor-valued surface conductivity, $\boldsymbol\sigma^\Sigma$, which may depend on coordinates $x,y$ and the frequency, $\omega$. Thus, allowing $\boldsymbol\sigma^\Sigma$ to act on vectors in $\mathbb{R}^3$, we use the matrix representation
\begin{equation*}
\boldsymbol\sigma^\Sigma=\begin{pmatrix} \sigma_{xx} & \sigma_{xy} & 0\\
                                 \sigma_{yx}  & \sigma_{yy} & 0\\
                                         0     &  0         & 0
                   \end{pmatrix}~,
\end{equation*}
where matrix elements $\sigma_{\ell m}$ ($\ell, m=x, y$) are in principle complex-valued functions of $x$, $y$ and $\omega$. (However, in Section~\ref{subsec:integral} $\boldsymbol\sigma^\Sigma$ is not allowed to depend on $y$). The ambient space has a constant dielectric permittivity, $\varepsilon$, and a constant magnetic permeability, $\mu$. For non-active 2D materials, the Hermitian part of this $\boldsymbol\sigma^\Sigma$ must be positive semidefinite; also, $\boldsymbol\sigma^\Sigma$ must obey the Onsager reciprocity relations.\cite{Onsager1931-I,Casimir1945} We note in passing that by causality $\boldsymbol\sigma^\Sigma(\omega)$ must be analytic in the upper $\omega$-plane (for $\Im\,\omega>0$), if $\omega$ becomes complex. \color{black}

The curl laws of the time harmonic Maxwell equations {\em outside the sheet $\Sigma$} read
\begin{equation}\label{eq:curl-laws}
\nabla\times \bfe=\iu\omega \bfb~,\ \nabla\times (\mu^{-1} \bfb)=-\iu\omega \varepsilon \bfe+\boldsymbol J_e\quad \mbox{in}\ \mathbb{R}^3\setminus\overline{\Sigma}~;\ \overline\Sigma:=\{(x,y,z): x\ge 0, z=0\}~.
\end{equation}
Here, $\bfe$ and $\bfb$ are the electric and magnetic fields, respectively; and $\boldsymbol J_e$ is the compactly supported current density of an external source. On $\Sigma$, we impose the boundary conditions~\cite{MML-JCP}
\begin{subequations}\label{eq:gen-bcs-Jind}
\begin{equation}\label{eq:gen-bcs}
\left[\ez\times \bfe\right]_\Sigma=0~,\quad \left[\ez\times \bfb\right]_\Sigma=\mu\,\bjs \quad \mbox{on}\ \Sigma~.
\end{equation}
Here, $[\boldsymbol Q]_\Sigma:=\boldsymbol Q(x,y,z=0^+)-\boldsymbol Q(x,y,z=0^-)$ for $x>0$, which denotes the jump of $\boldsymbol Q(x,y,z)$ across $\Sigma$; and $\bjs$ on $\Sigma$ is the vector-valued surface flux induced on the sheet, viz.,
\begin{equation}\label{eq:Jind}
\bjs(x,y)=\boldsymbol \sigma^\Sigma\, \bfe(x,y,z)=(\sigma_{xx}E_x+\sigma_{xy}E_y)\ex+(\sigma_{yx}E_x+\sigma_{yy}E_y)\ey\quad \mbox{for}\ z=0~,\ x>0~.
\end{equation}
\end{subequations}
We set $\bjs(x,y)\equiv 0$ if $x<0$. More generally, $\bjs$ can be a linear functional of $\bfe$ at $z=0$.\cite{MMSLL-preprint}

We alert the reader that~\eqref{eq:curl-laws} and~\eqref{eq:gen-bcs-Jind} introduce the volume current density,  $\boldsymbol J_e$, as {\em distinct} from the induced surface current density, $\bjs$, on $\Sigma$. This distinction is justified if the domain of Maxwell's equations is $\mathbb{R}^3\setminus\overline\Sigma$; and highlights the different physical origins of the two current densities. An alternate yet mathematically equivalent view, which we adopt for convenience in Section~\ref{subsec:integral}, is to extend the domain of Maxwell's laws to the whole Euclidean space and include $\bjs$ in their source term by treating it as a distribution (delta function in $z$).
\color{black}

We now discuss a suitable far-field condition.~\cite{King1963,Muller1969,TTWu1957} To determine the EP dispersion relation, we will set $\boldsymbol J_e=0$ and solve the ensuing homogeneous boundary value problem by assuming that the solution is a wave, the EP, that travels along and remains localized near the $y$-axis (Section~\ref{subsec:integral}). In this setting, the imposition of an outgoing wave in the $\pm z$-directions in addition to having an exponentially decaying \emph{and} outgoing wave in the $y$-direction may yield a solution that is exponentially increasing with $|z|$, similarly to the problem of the infinitely long microstrip.~\cite{TTWu1957} We will therefore consider solutions that {\em decay exponentially} in the directions perpendicular to the sheet. An implication of this assumption is outlined in the end of Section~\ref{sec:W-H}.

\subsection{Edge-plasmon polariton and integral equations for tangential electric field}
\label{subsec:integral}
Next, we derive integral equations for  $E_x$ and $E_y$ at $z=0$, by introducing the EP as a particular solution.
We assume that the surface conductivity, $\boldsymbol\sigma^\Sigma$, is independent of $y$ and the fields are traveling waves in the $y$-direction; the related wave number is to be determined. 

We invoke the vector potential, $\bas$, of the scattered field $(\boldsymbol E^{\rm sc}, \boldsymbol B^{\rm sc})$ in the Lorenz gauge.  The $\bas$ of course satisfies $\bfb^{\rm sc}=\nabla\times \bas$ outside $\overline\Sigma$. 
Our derivation of integral equations for the electric field here is akin to the derivation of the Pocklington integral equation for electric currents on thin cylindrical antennas in uniform media.~\cite{KingFikioris2002} Our integral formalism is a particular case of the ``electric field integral equation'' approach in electromagnetics.~\cite{Chew-book} 

To account for the EP, we consider fields of the form $\boldsymbol F(x,y,z)=e^{\iu q y} \widetilde {\boldsymbol F}(x,z;q)$ and replace $\nabla$ by 
$(\partial_x,\iu q,\partial_z)$, where $q$ is a complex wave number to be determined and $\boldsymbol F=\bfe,\,\bfb, \bas, \bjs, \boldsymbol J_e$. Now drop the tildes from all respective variables, which depend on $x$ or $z$, for ease of notation. This procedure amounts to taking the Fourier transform of Maxwell's equations and the boundary conditions with respect to $y$ (where $q$ is the `dual variable'). 

By~\eqref{eq:curl-laws} and~\eqref{eq:gen-bcs-Jind}, $\bas$ is due to the electron flow on the sheet and, thus, obeys the following nonhomogeneous Helmholtz equation on $\mathbb{R}^2$:
\begin{equation*}
(\Delta_{x,z}+k_0^2-q^2)\bas(x,z)=-\mu\, \bjs(x)\,\delta(z)\ \mbox{for\ all}\ (x,z)\ \mbox{in}\ \mathbb{R}^2~,
\end{equation*} 
where $\delta(z)$ is the Dirac delta function  and $\Delta_{x,z}$ is the 2D Laplacian ($\Delta_{x,z}=\partial^2/\partial x^2+\partial^2/\partial z^2$). The vector potential $\bas(x,z)$ is given in terms of the surface current, $\bjs(x)$, by~\cite{King1963}
\begin{subequations}\label{eqs:vec-pot}
\begin{equation}\label{eq:vec-pot-form}
\bas(x,z)=\mu \int_{\mathbb{R}} G(x,z;x',0)\, \bjs(x')\,\,\dxp~\quad \mbox{in}\ \mathbb{R}^2\setminus \overline\Sigma_2~,
\end{equation}
where $\overline\Sigma_2:=\{(x,z)\in\mathbb{R}^2: x\ge 0, z=0\}$ and $G(x,z;x',z')$ is given by 
\begin{equation}\label{eq:Green-fun}
G(x,z;x',z')=\textstyle{\frac{\iu}{4}}H_0^{(1)}\left(k_{\rm eff}\sqrt{(x-x')^2+(z-z')^2}\right)~,\quad k_{\rm eff}:=\sqrt{k_0^2-q^2}~,\ \Im\,k_{\rm eff}>0~,
\end{equation}
\end{subequations}
by use of the first-kind Hankel function, $H_0^{(1)}(w)$, of the zeroth order.\cite{Bateman-II}  
This $G$ comes from the (retarded) Green function for the scalar Helmholtz equation, with effective wave number $k_{\rm eff}$. Note that by~\eqref{eq:vec-pot-form} $\bas(x,z)$ is continuous everywhere,~\cite{Muller1969} and equals $\bas=A_x^{\rm sc} \ex+A_y^{\rm sc} \ey$ where each component $A_\ell^{\rm sc}=\mathbf e_\ell\cdot \bas$ is determined by $\mathbf e_\ell\cdot \bjs$ at $z=0$ ($\ell=x,\,y$). 
We compute
\begin{align*}
B_x^{\rm sc}(x,z)&=-\frac{\partial A_y^{\rm sc}}{\partial z}~,\quad B_y^{\rm sc}(x,z)=\frac{\partial A_x^{\rm sc}}{\partial z}~,\quad  B_z^{\rm sc}(x,z)=\frac{\partial A_y^{\rm sc}}{\partial x}-\iu q A_x^{\rm sc}\quad \mbox{in}\ \mathbb{R}^2\setminus\overline\Sigma_2~.
\end{align*}
Hence, by the Amp\`ere-Maxwell law from~\eqref{eq:curl-laws} we find the field components (defined in $\mathbb{R}^2\setminus\overline\Sigma_2$)
\begin{align*}
E_x^{\rm sc}(x,z)&=\frac{\iu\omega}{k_0^2}\left\{\left(\frac{\partial^2}{\partial x^2}+k_0^2\right)A_x^{\rm sc}+\iu q\frac{\partial A_y^{\rm sc}}{\partial x}\right\}~,\   
E_y^{\rm sc}(x,z)=\frac{\iu\omega}{k_0^2}\left(\iu q\frac{\partial A_x^{\rm sc}}{\partial x}+k_{\rm eff}^2 A_y^{\rm sc}\right)~,
\end{align*}
which are continuous across the half line $\Sigma_2:=\{(x,z)\in\mathbb{R}^2: x>0, z=0\}$, the projection of the physical sheet on the $xz$-plane. Thus, $E_x^{\rm sc}$ and $E_y^{\rm sc}$ obey the first condition in~\eqref{eq:gen-bcs}. One can verify that  ($\boldsymbol E^{\rm sc}$, $\boldsymbol B^{\rm sc}$) satisfies Faraday's law in~\eqref{eq:curl-laws} and the second condition in~\eqref{eq:gen-bcs}.

To obtain the desired integral equations for $E_x$ and $E_y$, we use~\eqref{eq:vec-pot-form}. Thus, we find
\begin{align*}
E_x^{\rm sc}(x,z)&=\frac{\iu\omega\mu}{k_0^2}\left\{\left(\frac{\partial^2}{\partial x^2}+k_0^2\right)\int_0^\infty G(x,z;x',0)\,\left[\sigma_{xx}E_x(x',0)+\sigma_{xy}E_y(x',0)\right]\,\dxp\right.\\
&\qquad \left. +\iu q\frac{\partial}{\partial x}\int_0^\infty G(x,z;x',0)\,\left[\sigma_{yx}E_x(x',0)+\sigma_{yy} E_y(x',0)\right]\,\dxp \right\}~,\\
E_y^{\rm sc}(x,z)&=\frac{\iu\omega\mu}{k_0^2}\left\{\iu q\frac{\partial}{\partial x} \int_0^\infty G(x,z;x',0)\,\left[\sigma_{xx}E_x(x',0)+\sigma_{xy}E_y(x',0)\right]\,\dxp \right. \\
&\qquad \left. +k_{\rm eff}^2 \int_0^\infty G(x,z;x',0)\,\left[\sigma_{yx}E_x(x',0)+\sigma_{yy}E_y(x',0)\right]\,\dxp\right\}\quad \mbox{in}\ \mathbb{R}^2\setminus\overline\Sigma_2~.
\end{align*}
In these expressions, we take the limit $z\to 0$ for $x\neq 0$.~\cite{Muller1969}  In the absence of any external source, when $\boldsymbol J_e\equiv 0$, we have $(E_x^{\rm sc}, E_y^{\rm sc})=(E_x, E_y)$. For ease of notation, define
\begin{equation}\label{eq:u-v-def}
 u(x):=E_x(x,0)~,\ v(x):=E_y(x,0)~,\ K(x;q):=G(x,0;0,0)=(\iu/4)H_0^{(1)}(k_{\rm eff}|x|)~.
\end{equation} 
The resulting system of (homogeneous) integral equations reads as 
\begin{subequations}\label{eqs:integral-rels-gen}
\begin{align}
u(x)&=\frac{\iu\omega\mu}{k_0^2}\left\{\left(\frac{{\rm d}^2}{\dx^2}+k_0^2\right)\int_0^\infty \dxp\, K(x-x';q)\,\left[\sigma_{xx}u(x')+\sigma_{xy}v(x')\right]\right. \notag \\
&\qquad \left. +\iu q\frac{{\rm d}}{\dx}\int_0^\infty \dxp\,K(x-x';q)\,\left[\sigma_{yx}u(x')+\sigma_{yy} v(x')\right] \right\}~, \label{eq:integral-rels-gen-Ex}\\
v(x)&=\frac{\iu\omega\mu}{k_0^2}\left\{\iu q\frac{{\rm d}}{\dx} \int_0^\infty \dxp\,K(x-x';q)\,\left[\sigma_{xx}u(x')+\sigma_{xy}v(x')\right] \right. \notag \\
&\qquad \left. +k_{\rm eff}^2 \int_0^\infty \dxp\, K(x-x';q)\,\left[\sigma_{yx}u(x')+\sigma_{yy}v(x')\right]\right\}~,\quad  \mbox{all}\ x\ \mbox{in}\  \mathbb{R}\setminus\{0\}~. \label{eq:integral-rels-gen-Ey}
\end{align}
\end{subequations}
Let us now formally extend the domain of these equations to the whole $\mathbb{R}$. We have
\begin{subequations}\label{eqs:IE-anis-conductivity}
\begin{equation}\label{eq:integral-rels-gen-matr}
\begin{pmatrix} u(x) \\
                v(x) 
\end{pmatrix} =    \frac{\iu\omega\mu}{k_0^2}
\begin{pmatrix}
\displaystyle \frac{{\rm d}^2}{\dx^2}+k_0^2 &  \quad \displaystyle \iu q \frac{{\rm d}}{\dx} \\
  \displaystyle \iu q\frac{{\rm d}}{\dx}      &   k_{\rm eff}^2   	
\end{pmatrix} 
\int_0^\infty \dxp\, K(x-x';q)\,\boldsymbol\sigma^\Sigma_2\,  
\begin{pmatrix} u(x') \\
                v(x') 
\end{pmatrix} \quad  \mbox{all}\ x\ \mbox{in}\ \mathbb{R}~,            	
\end{equation}
where
\begin{equation}\label{eq:sigma2}
\boldsymbol\sigma_2^\Sigma:=
\begin{pmatrix}
	\sigma_{xx} & \sigma_{xy} \\
	\sigma_{yx} & \sigma_{yy}
\end{pmatrix}~.	
\end{equation}
\end{subequations}
Note that this $\boldsymbol\sigma_2^\Sigma$ may depend on the coordinate $x'$. For a generalization of~\eqref{eq:integral-rels-gen-matr} to 2D materials in which the induced surface current density is a linear functional of $(u(x), v(x))$, \color{black} see Ref.~\onlinecite{MMSLL-preprint}. The problem is to find $q$ so that matrix equation~\eqref{eq:integral-rels-gen-matr} admits nontrivial solutions.

The right-hand side of~\eqref{eq:integral-rels-gen-matr} involves the field components tangential to the physical sheet, for $x'>0$ under the integral sign. Strictly speaking, \eqref{eq:integral-rels-gen-matr} yields a system of integral equations for $u(x)$ and $v(x)$ by restriction of both sides of this equation to $x>0$. By solving these equations for an isotropic sheet, we will verify that any nontrivial, admissible electric field component $u(x)$, which is normal to the edge, is singular and discontinuous at $x=0$  (Section~\ref{sec:solution}). In contrast, $v(x)$ turns out to be continuous at $x=0$ (Section~\ref{sec:solution}).

We can state a more precise definition of the EP; cf. Refs.~\onlinecite{Fetter1985,VolkovMikhailov1988}.
\medskip

{\em  Definition 1 (Edge plasmon-polariton). The {\rm EP} amounts to nontrivial integrable solutions $(u,v)$ and corresponding wave number, $q$, of~\eqref{eq:integral-rels-gen-matr} ($u,\,v\in L^1(\mathbb{R})$). The {\rm EP dispersion relation} describes how the $q$ of this solution is related to the angular frequency, $\omega$.}
\medskip

An assumption underlying Definition~1 is that nontrivial integrable solutions $u(x)$ and $v(x)$ of~\eqref{eq:integral-rels-gen-matr}, and the corresponding $q$'s, exist for some range of frequencies $\omega$, given some meaningful model for the surface conductivity 
$\boldsymbol\sigma_2^\Sigma$. We will construct such solutions by the Wiener-Hopf method for the simplified model with $\boldsymbol\sigma_2^\Sigma=\sigma \boldsymbol I_2$ where $\boldsymbol I_2$ is the $2\times 2$ unit matrix and $\sigma$ is a spatially constant but $\omega$-dependent scalar quantity (Sections~\ref{sec:W-H} and~\ref{sec:solution}).

\section{Homogeneous sheet: Coupled functional equations}
\label{sec:W-H}
In this section, we reduce~\eqref{eq:integral-rels-gen-matr} to a system of functional equations on the real line for Fourier-transformed fields via the Wiener-Hopf method,~\cite{Krein1962,Masujima-book}
if
\begin{equation*}
\sigma_{xx}~,\ \sigma_{xy}~,\ \sigma_{yx}~,\ \sigma_{yy}\ \mbox{are\ spatially\ constant}.	\end{equation*}

Equation~\eqref{eq:integral-rels-gen-matr} is recast to the system
\begin{subequations}\label{eq:integral-const}
\begin{equation}\label{eq:integral-rels-const}
\begin{pmatrix} u(x) \\
                v(x) 
\end{pmatrix} =    \frac{\iu\omega\mu}{k_0^2}
\begin{pmatrix}
\displaystyle \frac{{\rm d}^2}{\dx^2}+k_0^2 &  \quad \displaystyle \iu q \frac{{\rm d}}{\dx} \\
  \displaystyle \iu q\frac{{\rm d}}{\dx}      &   k_{\rm eff}^2   	
\end{pmatrix} 
\boldsymbol\sigma^\Sigma_2\,\int_{-\infty}^\infty \dxp\, K(x-x';q)\,  
\begin{pmatrix} u_{>}(x') \\
                v_{>}(x') 
\end{pmatrix} \quad  x\ \mbox{in}\ \mathbb{R}~,            	
\end{equation}
where
\begin{equation}\label{eq:integral-const-defs}
\begin{pmatrix} u_{>}(x) \\
                v_{>}(x) 
\end{pmatrix}
:=	
\begin{pmatrix} u(x) \\
                v(x) 
\end{pmatrix}
\quad \mbox{if}\ x>0~,\quad 
\begin{pmatrix} u_{>}(x) \\
                v_{>}(x) 
\end{pmatrix}
\equiv 0\quad  \mbox{if}\ x<0~.
\end{equation}
\end{subequations}
Furthermore, we define the functions $u_{<}(x)$ and $v_{<}(x)$ via the relations
$u(x)=u_{>}(x)+u_{<}(x)$ and $v(x)=v_{>}(x)+v_{<}(x)$ for all $x$ in $\mathbb{R}\setminus \{0\}$.

The Fourier transform of $f(x)$, with $f\in L^1(\mathbb{R})$, is
\begin{equation*}
\widehat f(\xi)=	\int_{-\infty}^{\infty} \dx\, f(x) e^{-\iu\xi x}~.
\end{equation*}
Hence, if $f(x)\equiv 0$ for $x<0$ then $\widehat f(\xi)$ is analytic in the lower half $\xi$-plane, $\mathbb{C}_-$; whereas $\widehat f(\xi)$ is analytic in the upper half plane, $\mathbb{C}_+$, if $f(x)\equiv 0$ for $x>0$.~\cite{Krein1962,PaleyWiener,Wiener-book}

The application of the Fourier transform in $x$ to~\eqref{eq:integral-rels-const} yields the system
\begin{subequations}\label{eq:integral-rels-FT-3}
\begin{equation}\label{eq:integral-rels-FT}
\begin{pmatrix} \widehat u_+(\xi) \\
                \widehat v_+(\xi) 
\end{pmatrix} +
\begin{pmatrix} \widehat u_-(\xi) \\
                \widehat v_-(\xi) 
\end{pmatrix}
=    \frac{\iu\omega\mu}{k_0^2}\, \widehat K(\xi;q)
\begin{pmatrix}
\displaystyle k_0^2-\xi^2 &  \ \displaystyle (\iu q)(\iu\xi)  \\
  \displaystyle (\iu q)(\iu\xi)      &   \ k_0^2-q^2   	
\end{pmatrix} 
\boldsymbol\sigma^\Sigma_2\, 
\begin{pmatrix} \widehat u_{-}(\xi) \\
                \widehat v_{-}(\xi) 
\end{pmatrix}\quad \mbox{all}\ \xi\ \mbox{in}\ \mathbb{R}~,            	
\end{equation}
which describes two coupled functional equations on the real line. 
In the above, we have  
\begin{equation}\label{eq:kernel-FT}
\widehat K(\xi;q)=\int_{-\infty}^\infty \dx \,K(x;q) \, e^{-\iu\xi x}=\frac{\iu}{2}(k_{\rm eff}^2-\xi^2)^{-1/2}~,\quad k_{\rm eff}=\sqrt{k_0^2-q^2}~,
\end{equation}
where $\Im\sqrt{k_{\rm eff}^2-\xi^2}>0$ since we impose decay of $\widehat G(\xi,z;0,0)$ with $\Im\,k_{\rm eff}>0$ as $|z|\to\infty$; see~\eqref{eq:Green-fun}.  With this choice of the top Riemann sheet $\widehat K(\xi;q)$ is an even function of $\xi$.  Note that the requisite branch cuts, which emanate from $\xi=\pm k_{\rm eff}=\pm \sqrt{k_0^2-q^2}$ ($\Im\,k_{\rm eff}>0$), lie in $\mathbb{C}_\pm$ and are infinite and symmetric with respect to the origin. We also define 
\begin{equation}\label{eq:+-fcns-def}
\begin{pmatrix} \widehat u_{\pm}(\xi) \\
                \widehat v_{\pm}(\xi) 
\end{pmatrix}
= \int_{-\infty}^{\infty}\dx\ 
\begin{pmatrix} 
 u_{< \atop >}(x) \\
 v_{< \atop >}(x) 
\end{pmatrix}
e^{-\iu\xi x}~.
\end{equation}
\end{subequations}
Of course, $\widehat u_\pm(\xi)$ and $\widehat v_\pm(\xi)$ depend on $q$; for ease of notation, we suppress this dependence.

Two comments are in order. First, $\widehat u_\pm(\xi)$ and $\widehat v_\pm(\xi)$ for real $\xi$ are viewed as limits of the corresponding analytic functions as $\xi$ approaches the real axis from $\mathbb{C}_+$ or $\mathbb{C}_-$. Thus, \eqref{eq:integral-rels-FT} expresses a Riemann-Hilbert problem on the real line. This type of problem, and the respective matrix Wiener-Hopf integral equation associated with it, can be solved explicitly, with the solution in simple closed form, only in a limited number of cases; see, e.g.,  Refs.~\onlinecite{GohbergKrein1960,WuWu1963,Abrahams1997}. We will solve~\eqref{eq:integral-rels-FT} explicitly for the special case with a scalar constant conductivity, i.e., if $\boldsymbol\sigma_2^\Sigma=\sigma \boldsymbol I_2$ where $\sigma$ is a scalar constant in $x$ and $y$ and $\boldsymbol I_2={\rm diag}(1,1)$ (Section~\ref{sec:solution}). Second, recall that we impose $\Im \sqrt{k_{\rm eff}^2-\xi^2}>0$ with $\Im\,k_{\rm eff}>0$ in the $\xi$-plane. Suppose for a moment that $\Re\,q>0$ and $\Im\,q>0$, i.e., the EP is an outgoing and decaying wave in the positive $y$-direction; then, $\Im\,k_{\rm eff}^2=\Im (k_0^2-q^2)<0$ if the ambient medium is lossless ($k_0>0$). Hence, the condition $\Re\,k_{\rm eff}<0$ must be satisfied, given that $\Im\,k_{\rm eff}>0$. By the prescribed choice of the branch cut for $\sqrt{k_{\rm eff}^2-\xi^2}$ and the respective integration path in the $\xi$-plane, we conclude that $\Re\sqrt{k_{\rm eff}^2-\xi^2}<0$; cf. Ref.~\onlinecite{TTWu1957}. The sign reversal of $\Re\,q$, i.e., the mapping $q\mapsto -q^*$, causes the sign change of $\Re\sqrt{k_{\rm eff}^2-\xi^2}$.

\section{Edge plasmon on isotropic homogeneous sheet}
\label{sec:solution}
In this section, we restrict attention to the case with an isotropic and homogeneous conducting sheet. Hence, we set
\begin{equation*}
\boldsymbol \sigma_2^\Sigma=\sigma\,\boldsymbol I_2~,	
\end{equation*}
where $\sigma$ is a scalar function of $\omega$ with $\Re\,\sigma(\omega)\ge 0$. We will explicitly solve~\eqref{eq:integral-rels-FT} via a suitable transformation of $(\widehat u_\pm(\xi), \widehat v_\pm(\xi))$ and subsequent factorizations in the $\xi$-plane.~\cite{Krein1962,GohbergKrein1960}

Equation~\eqref{eq:integral-rels-FT} is recast to the system 
\begin{equation*}
\begin{pmatrix} \widehat u_+(\xi) \\
                \widehat v_+(\xi) 
\end{pmatrix} +
\begin{pmatrix} \widehat u_-(\xi) \\
                \widehat v_-(\xi) 
\end{pmatrix}
=    \frac{\iu\omega\mu\sigma}{k_0^2}\, \widehat K(\xi;q)
\begin{pmatrix}
\displaystyle k_0^2-\xi^2 &  \ \displaystyle (\iu q)(\iu\xi)  \\
  \displaystyle (\iu q)(\iu \xi)      &   \ k_0^2-q^2   	
\end{pmatrix}  
\begin{pmatrix} \widehat u_{-}(\xi) \\
                \widehat v_{-}(\xi) 
\end{pmatrix}\qquad (\mbox{all\ real}\ \xi)~.          	\end{equation*}
Now define the matrix 
\begin{subequations}
\begin{equation}\label{eq:K-matrix-def}
\bkm(\xi;q):= \begin{pmatrix}
\displaystyle 1-\frac{\iu\omega\mu\sigma}{k_0^2}(k_0^2-\xi^2)\widehat K(\xi;q) &  \ \displaystyle -\frac{\iu\omega\mu\sigma}{k_0^2}(\iu q)(\iu \xi) \widehat K(\xi;q)  \\
  \displaystyle -\frac{\iu\omega\mu\sigma}{k_0^2}(\iu q)(\iu\xi)\widehat K(\xi; q)      &   \ \displaystyle 1-\frac{\iu\omega\mu\sigma}{k_0^2}(k_0^2-q^2) \widehat K(\xi;q)   	\end{pmatrix}~.  
\end{equation}
Accordingly, the functional equations under consideration are expressed by 
\begin{equation}\label{eq:integral-rels-FT-const}
\bkm(\xi; q)
\begin{pmatrix} \widehat u_{-}(\xi) \\
                \widehat v_{-}(\xi) 
\end{pmatrix}
+
\begin{pmatrix} \widehat u_{+}(\xi) \\
                \widehat v_{+}(\xi) 
\end{pmatrix}  
=0\qquad (\mbox{all\ real}\ \xi)~.          	
\end{equation}
\end{subequations}

\subsection{Linear transformation and explicit solution}
\label{subsec:explicit}
The key observation is that in the present setting we can explicitly define a matrix valued function, $\boldsymbol{\mathcal T}(\xi;q)$, such that the transformed vector valued function 
\begin{equation}\label{eq:lin-transf}
\begin{pmatrix}
U(\xi) \\
V(\xi)
\end{pmatrix}
:= \boldsymbol{\mathcal T}(\xi;q) 
\begin{pmatrix}
\widehat u(\xi)\\
\widehat v(\xi)
\end{pmatrix}
\end{equation} 
has the following properties: (i) ($U_s(\xi),  V_s(\xi))^T=\boldsymbol{\mathcal T}(\xi;q) (\widehat u_s(\xi),\widehat v_s(\xi))^T$ where the superscript $T$ denotes transposition and $s=\pm$; and (ii) the components $U_s(\xi)$ and $V_s(\xi)$ separately satisfy two (decoupled) functional equations on the real axis. The first property [item (i)] directly follows from \eqref{eq:lin-transf} if each matrix element in $\boldsymbol{\mathcal T}(\xi;q)$ is an entire function of $\xi$.

To this end, we diagonalize $\bkm(\xi;q)$. Consider an invertible matrix $\bsm$ such that
\begin{equation*}
\bkm(\xi;q)=\bsm(\xi;q)\, {\rm diag}(\mathcal P_1(\xi;q),\mathcal P_2(\xi;q))\,\bsm^{-1}(\xi;q)\qquad (\xi\in\mathbb{C})~,
\end{equation*}
where $\mathcal P_j(\xi;q)$ are eigenvalues of $\bkm(\xi;q)$ ($j=1,\,2$). The associated eigenvalues satisfy 
\begin{align*}
&\mathcal P^2-\left\{2-\frac{i\omega\mu\sigma}{k_0^2}(2k_0^2-q^2-\xi^2)\widehat K(\xi;q)\right\}\mathcal P+1-\frac{i\omega\mu\sigma}{k_0^2}(2k_0^2-q^2-\xi^2)\widehat K(\xi;q)\\
&\qquad + \left(\frac{\iu \omega\mu\sigma}{k_0}\right)^2 (k_{\rm eff}^2-\xi^2) \,\widehat K(\xi;q)^2=0~,
\end{align*}
which has the distinct solutions
\begin{equation}\label{eq:K-eigenv}
\mathcal P_1=\mathcal P_{\rm TM}(\xi;q):=1-\frac{\iu\omega\mu\sigma}{k_0^2}(k_{\rm eff}^2-\xi^2)\widehat K(\xi; q)~,\quad \mathcal P_2=\mathcal P_{\rm TE}(\xi;q):=1-\iu\omega\mu\sigma\widehat K(\xi; q)
\end{equation}
where $k_{\rm eff}=\sqrt{k_0^2-q^2}$ and $\widehat K(\xi;q)$ is defined by~\eqref{eq:kernel-FT}. Note that $\mathcal P_{\rm TM}(\pm \iu q;q)=\mathcal P_{\rm TE}(\pm\iu q;q)$. 

We will show how the contributions from $\mathcal P_{\rm TM}$ and $\mathcal P_{\rm TE}$ enter the EP dispersion relation (Section~\ref{subsec:EP-disp}). In regard to these eigenvalues, $\mathcal P_{\rm TM}$ corresponds to TM polarization while $\mathcal P_{\rm TE}$ amounts to TE polarization. This terminology is motivated as follows: The roots $\xi$ of $\mathcal P_{\rm TM}(\xi;0)=0$ or $\mathcal P_{\rm TE}(\xi;0)=0$ provide the propagation constants in the $x$-direction for the TM- or TE-polarized SP on the respective infinite 2D conducting material with $q=0$.\cite{Hanson2008,Bludov2013,ML2016} Alternatively, by replacing $\xi$ by $\sqrt{q_x^2+q_y^2}$ in these roots, where $q_\ell$ is the wave number in the $\ell$-direction ($\ell=x,y$), and solving for $\omega(q_x,q_y)$ one recovers the continuum energy spectrum of the TM- or TE-polarized SP on the infinite sheet.\cite{Bludov2013,Pitarkeetal2007} The roots $\xi$ for each case are present in the top Riemann sheet under suitable conditions on the phase of $\sigma$ 
(see Section~\ref{subsec:far-f}).\cite{Hanson2008,Bludov2013,ML2016,MML2017}

By an elementary calculation, eigenvectors of $\bkm(\xi;q)$ are given by
\begin{equation*}
\begin{pmatrix}
\iu \xi \cr
\iu q
\end{pmatrix}
\ \mbox{for}\ \mathcal P=\mathcal P_{\rm TM}\quad \mbox{and}\quad
\begin{pmatrix}
\iu q \cr
-\iu \xi
\end{pmatrix}
\ \mbox{for}\ \mathcal P=\mathcal P_{\rm TE}~,
\end{equation*}
which depend on the material parameters through $q$ if the latter satisfies a dispersion relation.
Hence, the matrix $\boldsymbol{\mathcal S}$ can be taken to be equal to
\begin{equation}\label{eq:S-matrix-def}
\boldsymbol{\mathcal S}(\xi;q)=
\begin{pmatrix}
\iu \xi &  \iu q \cr
\iu q    & -\iu \xi
\end{pmatrix}~,
\end{equation}
which is an entire matrix valued function of $\xi$, and invertible for all complex $\xi$ with $\xi\neq \pm \iu q$.
Once we compute $\boldsymbol{\mathcal S}^{-1}=-(\xi^2+q^2)^{-1}\boldsymbol{\mathcal S}$, we write
\begin{equation*}
\bkm(\xi;q)=-\frac{1}{q^2+\xi^2}
\boldsymbol{\mathcal S}(\xi;q)
\begin{pmatrix}
\mathcal P_{\rm TM}(\xi; q) & 0 \cr
0  &  \mathcal P_{\rm TE}(\xi;q)
\end{pmatrix}
\boldsymbol{\mathcal S}(\xi;q)~.
\end{equation*}
Accordingly, by~\eqref{eq:integral-rels-FT-const} we obtain the expression
\begin{equation*}
\begin{pmatrix}
\mathcal P_{\rm TM}(\xi; q) & 0 \cr
0  &  \mathcal P_{\rm TE}(\xi;q)
\end{pmatrix}
\boldsymbol{\mathcal S}(\xi;q)
\begin{pmatrix}
\widehat u_-(\xi)\cr
\widehat v_-(\xi)
\end{pmatrix}
+
\boldsymbol{\mathcal S}(\xi;q)
\begin{pmatrix}
\widehat u_+(\xi)\cr
\widehat v_+(\xi)
\end{pmatrix}
=0\quad (\mbox{all\ real}\ \xi)~.
\end{equation*}
Thus, by recourse to~\eqref{eq:lin-transf} we can set
\begin{equation*}
\boldsymbol{\mathcal T}(\xi;q)=\boldsymbol{\mathcal S}(\xi;q)=
\begin{pmatrix}
\iu \xi &     \iu q \cr
\iu q   &      -\iu \xi 
\end{pmatrix}~.
\end{equation*}
This choice implies the transformation $(\widehat u,\widehat v)\mapsto (U, V)$ with
\begin{equation}\label{eq:UV-transf-def}
U(\xi)=\iu \xi\, \widehat u(\xi)+\iu q\, \widehat v(\xi)~,\quad 
V(\xi)=\iu q\,\widehat u(\xi)-\iu \xi\,\widehat v(\xi)~.
\end{equation}
Evidently, $(U_\pm(\xi), V_\pm(\xi))^T$ may result from the 
application of  $\boldsymbol{\mathcal T}(\xi;q)$ to $(\widehat u_\pm(\xi), \widehat v_\pm(\xi))^T$. 
\medskip

{\em Remark~1 (On the transformation for $\widehat u$ and $\widehat v$). Equations~\eqref{eq:UV-transf-def} represent the Fourier tranforms with respect to $x$ of $-\partial E_z(x,z)/\partial z=\partial E_x/\partial x+\iu q\, E_y$ and $-i\omega B_z(x,z)=\iu q E_x(x,z)-\partial E_y/\partial x$ at $z=0$ by omission of any boundary terms for $E_x(x,0)$ and $E_y(x,0)$ (as $x\to 0$). This absence of boundary terms is consistent with the presence of a non-integrable singularity of $\partial E_x/\partial x$ and the continuity of $E_y(x,z)$ as $x\to 0$ at $z=0$ (Section~\ref{subsec:EP_and_SP}). }
\medskip 

We return to the task of computing $\widehat{u}(\xi)$ and $\widehat{v}(\xi)$. The functions $U_\pm(\xi)$ and $V_\pm(\xi)$ obey
\begin{subequations}\label{eq:transf-func-eqs}
\begin{align}
&\mathcal P_{\rm TM}(\xi;q)U_-(\xi)+U_+(\xi)=0~,\label{eq:func-eq-TM}\\
&\mathcal P_{\rm TE}(\xi;q)V_-(\xi)+V_+(\xi)=0\quad \mbox{for\ all}\ \xi\ \mbox{in}\ \mathbb{R}~.\label{eq:func-eq-TE}
\end{align}
\end{subequations}
Hence, loosely speaking, the contributions from the TE- and TM-polarizations are now decoupled.
Our goal is to solve~\eqref{eq:transf-func-eqs} explicitly (Section~\ref{subsec:EP-disp}); and then account for transformation~\eqref{eq:UV-transf-def} in order to obtain $q$ as well as the corresponding nontrivial $\widehat u_\pm(\xi)$ and $\widehat v_\pm(\xi)$.

We should alert the reader that the approach of matrix diagonalization, which we apply above, is tailored to the present {\em isotropic} model of the surface conductivity. This approach  
is in principle not suitable for a strictly anisotropic conductivity in functional equations~\eqref{eq:integral-rels-FT-3}. This limitation can be attributed to the ensuing analytic structure of the matrix $\boldsymbol{\mathcal S}(\xi;q)$. 


\subsection{Derivation of EP dispersion relation}
\label{subsec:EP-disp}
Let us assume that for all admissible $q$ the functions $\mathcal P_{\rm TM}(\xi;q)$ and $\mathcal P_{\rm TE}(\xi;q)$ satisfy
\begin{equation*}
\mathcal P_{\rm TM}(\xi;q)\neq 0\ \mbox{and}\ \mathcal P_{\rm TE}(\xi;q)\neq 0\quad \mbox{for\ all\ real}\ \xi~.
\end{equation*}
Hence, the functions $\ln\mathcal P_{\rm TM}(\xi)$ and $\ln\mathcal P_{\rm TE}(\xi)$, which we invoke below, are analytic in a vicinity of the real axis in the $\xi$-plane. \color{black} 
The above conditions imply that the respective bulk SPs, for fixed $q$, do not have real propagation constants in the first Riemann sheet ($\Im\sqrt{k_0^2-q^2-\xi^2}>0$).
To simplify the notation, we henceforth suppress the $q$-dependence in quantities such as $\mathcal P_{\rm TM}$ and $\mathcal P_{\rm TE}$.

In order to solve~\eqref{eq:transf-func-eqs} we need to carry out factorizations of $\mathcal P_{\rm TM}(\xi)$ and $\mathcal P_{\rm TE}(\xi)$, i.e., determine `split functions' $Q_\pm(\xi)$ and $R_\pm(\xi)$ such that~\cite{Krein1962}
\begin{equation}\label{eq:QR-split-def}
Q(\xi):=\ln\mathcal P_{\rm TM}(\xi)=Q_+(\xi)+Q_-(\xi)~,\quad R(\xi):=\ln\mathcal P_{\rm TE}(\xi)=R_+(\xi)+R_-(\xi)~,
\end{equation}
which is a classic problem in complex analysis. 
The EP dispersion relation will be expressed in terms of functions $Q_\pm$ and $R_\pm$. Note that $Q(\xi)$ and $R(\xi)$ are even functions in the top Riemann sheet.

It is useful to introduce the (vector-valued) {\em index}, $\boldsymbol\nu$, for functional equations~\eqref{eq:transf-func-eqs}. This $\boldsymbol\nu$ expresses the indices associated with $\mathcal P_{\rm TM}(\xi)$ and $\mathcal P_{\rm TE}(\xi)$ on the real axis, viz.,~\cite{Krein1962,Masujima-book}
\begin{equation*}
\boldsymbol\nu:=\frac{1}{2\pi\iu}\lim_{M\to +\infty}\int_{-M}^M  
\begin{pmatrix}
\{{\mathcal P'_{\rm TM}}(\xi)/\mathcal P_{\rm TM}(\xi)\} \cr
\{{\mathcal P'_{\rm TE}}(\xi)/\mathcal P_{\rm TE}(\xi)\} 
\end{pmatrix} \,\dxi=
\frac{1}{2\pi}\lim_{M\to +\infty}
\begin{pmatrix}
\arg\mathcal P_{\rm TM}(\xi) \cr
\arg\mathcal P_{\rm TE}(\xi) 
\end{pmatrix}\Biggl|_{\xi=-M}^M~,
\end{equation*}
where the prime here denotes differentiation with respect to the Fourier variable $\xi$. The components of this $\boldsymbol\nu$ express the changes of the values for $(2\pi\iu)^{-1}\ln \mathcal P_{\rm TM}(\xi)$ and $(2\pi\iu)^{-1}\ln\mathcal P_{\rm TE}(\xi)$ as $\xi$ moves between the extremities of the real axis. Thus, each component of $\boldsymbol\nu$ is the winding number with respect to the origin of a contour, $C_0^{\varpi}$, in the complex $\mathcal P_\varpi$-plane under the mapping $\xi\mapsto \mathcal P_\varpi(\xi)$ which maps the real axis to $C_0^\varpi$  ($\varpi={\rm TM}$ or ${\rm TE}$).

Because $\mathcal P_{\rm TM}(\xi)$ and $\mathcal P_{\rm TE}(\xi)$ are even functions of $\xi$, we can assert that
\begin{equation}\label{eq:index-zero}
\boldsymbol\nu=0
\end{equation}
which implies that splitting~\eqref{eq:QR-split-def} makes sense and can be carried out directly via the Cauchy integral formula.\cite{Krein1962,MML2017}  \color{black} 
In contrast, for certain {\em strictly anisotropic} conducting sheets, 
 the index for the underlying Wiener-Hopf integral equations in the quasi-electrostatic approach may be nonzero, which implies distinct possibilities regarding the existence, or lack thereof, of the  EP.~\cite{VolkovMikhailov1988,MMSLL-preprint} This material anisotropy lies beyond the scope of the present paper.

Therefore, we can directly apply the Cauchy integral formula and obtain~\cite{Krein1962,MML2017}
\begin{subequations}\label{eq:QR-def}
\begin{align}
Q_\pm(\xi)&=\pm \frac{1}{2\pi \iu}\int_{-\infty}^\infty \frac{Q(\xi')}{\xi'-\xi}\ \dxip=\pm \frac{\xi}{\iu\pi}\int_0^\infty \frac{Q(\xi')}{{\xi'}^2-\xi^2}\ \dxip~, \label{eq:Q-int-def}\\
R_\pm(\xi)&=\pm \frac{1}{2\pi \iu}\int_{-\infty}^\infty\frac{R(\xi')}{\xi'-\xi}\ \dxip=\pm \frac{\xi}{\iu\pi}\int_{-\infty}^\infty \frac{R(\xi')}{{\xi'}^2-\xi^2}\ \dxip\quad (\pm\Im\,\xi>0)~, \label{eq:R-int-def}
\end{align}
\end{subequations}
in view of definitions~\eqref{eq:QR-split-def}.
Equations~\eqref{eq:transf-func-eqs} then read
\begin{equation*}
e^{Q_-(\xi)}U_-(\xi)=-e^{-Q_+(\xi)}U_+(\xi)~,\quad e^{R_-(\xi)}V_-(\xi)=-e^{-R_+(\xi)}V_+(\xi)\qquad \mbox{for\ all}\ \xi\ \mbox{in}\ \mathbb{R}~.
\end{equation*}
By analytic continuation of each side of the above equations to complex $\xi$, in $\mathbb{C}_+$ or $\mathbb{C}_-$, we infer that there exist entire functions $\mathcal E_j(\xi)$ ($j=1,\,2$) such
that~\cite{Krein1962}
\begin{subequations} \label{eq:UV-soln-E}
\begin{align}
& e^{Q_-(\xi)}U_-(\xi)=-e^{-Q_+(\xi)}U_+(\xi)=\mathcal E_1(\xi)~,\\
& e^{R_-(\xi)}V_-(\xi)=-e^{-R_+(\xi)}V_+(\xi)=\mathcal E_2(\xi)\qquad \mbox{for\ all}\ \xi\ \mbox{in}\ \mathbb{C}~.
\end{align}
\end{subequations}
Each of these $\mathcal E_j(\xi)$ can be determined by examination of $Q_\pm(\xi)$, $R_\pm(\xi)$, $U_\pm(\xi)$ and $V_\pm(\xi)$ 
as $\xi\to\infty$ in $\mathbb{C}_+$ or $\mathbb{C}_-$. It is compelling to consider only polynomials as candidates for $\mathcal E_j(\xi)$.

Let us now discuss in detail the issue of determining $\mathcal E_j(\xi)$. Recall that the electric-field components 
$E_x(x,0)$ and $E_y(x,0)$ are assumed to be integrable on $\mathbb{R}$. Hence, $\widehat u_\pm(\xi)\to 0$ and $\widehat v_\pm(\xi)\to 0$ as $\xi\to\infty$ in $\mathbb{C}_\pm$.~\cite{PaleyWiener,Wiener-book} By 
transformation~\eqref{eq:UV-transf-def}, we infer that
\begin{equation*}
U_\pm(\xi),\,V_\pm(\xi)\ \mbox{cannot\ grow\ as\ fast\ as}\ \xi
\end{equation*}
in the limit $\xi\to\infty$ in $\mathbb{C}_\pm$. To express this behavior, we write $|U_\pm(\xi)|< \mathcal O(\xi)$ and $|V_\pm(\xi)|<\mathcal O(\xi)$ as $\xi\to\infty$ in $\mathbb{C}_\pm$. Now consider the asymptotics for $Q_\pm(\xi)$ and $R_\pm(\xi)$ when $|\xi|$ is large; see the Appendix. We can assert that
\begin{equation*}
e^{Q_\pm(\xi)}=\mathcal O(\sqrt{\xi})\ \mbox{and}\ e^{R_\pm(\xi)}\to 1\quad \mbox{as}\ \xi\to\infty\ \mbox{in}\ \mathbb{C}_\pm~.
\end{equation*}
These estimates imply that
\begin{equation*}
|e^{Q_-(\xi)}U_-(\xi)|< \mathcal O(\xi\sqrt{\xi})\ \mbox{and}\ 
|e^{-Q_+(\xi)}U_+(\xi)|<\mathcal O(\sqrt{\xi})\quad \mbox{as}\ \xi\to\infty
\end{equation*}
in $\mathbb{C}_-$ or $\mathbb{C}_+$, respectively. In a similar vein, we have
\begin{equation*}
|e^{\mp R_\pm(\xi)}V_\pm(\xi)|<\mathcal O(\xi)\quad \mbox{as}\ \xi\to\infty\ \mbox{in}\ \mathbb{C}_\pm~.
\end{equation*}

Hence, we find that the entire functions $\mathcal E_1(\xi)$ and $\mathcal E_2(\xi)$ satisfy
\begin{equation*}
\mathcal E_1(\xi)<\mathcal O(\sqrt{\xi})\ \mbox{and}\ \mathcal E_2(\xi)<\mathcal O(\xi)\ \mbox{as}\ \xi\to\infty\ \mbox{in}\ \mathbb{C}~.
\end{equation*}
Thus, resorting to Liouville's theorem, we conclude that
\begin{equation}\label{eq:E-const}
\mathcal E_1(\xi)=C_1={\rm const}.\quad \mbox{and}\quad \mathcal E_2(\xi)=C_2={\rm const}.\qquad \mbox{for\ all}\ \xi\in\mathbb{C}~.
\end{equation}
These constants, $C_1$ and $C_2$, have units of electric field and are both arbitrary so far.

We proceed to determine $\widehat u_\pm(\xi)$ and $\widehat v_\pm(\xi)$ in terms of $C_1$ and $C_2$, and then obtain the EP dispersion relation. Equations~\eqref{eq:UV-soln-E} and~\eqref{eq:E-const} lead to
\begin{equation*}
U_\pm(\xi)=\mp C_1 e^{\pm Q_\pm(\xi)}~,\quad V_\pm(\xi)=\mp C_2 e^{\pm R_\pm(\xi)}~.
\end{equation*}
In view of transformation~\eqref{eq:UV-transf-def}, we readily obtain the formulas
\begin{subequations}\label{eq:uv-soln-minus}
\begin{align}
\widehat u_-(\xi)&=-\frac{\iu \xi\, U_-(\xi)+\iu q\, V_-(\xi)}{q^2+\xi^2}=-\frac{\iu \xi\,C_1 e^{-Q_-(\xi)}+\iu q\,C_2 e^{-R_-(\xi)}}{q^2+\xi^2}~,\label{eq:u-soln-minus}\\
\widehat v_-(\xi)&=-\frac{\iu q\,U_-(\xi)-\iu \xi\,V_-(\xi)}{q^2+\xi^2}=-\frac{\iu q\,C_1 e^{-Q_-(\xi)}-\iu \xi\,C_2e^{-R_-(\xi)}}{q^2+\xi^2}~, \label{eq:v-soln-minus}
\end{align}
\end{subequations}
for the fields $u_>(x)$ and $v_>(x)$, along with the formulas
\begin{subequations}\label{eq:uv-soln-plus}
\begin{align}
\widehat u_+(\xi)&=-\frac{\iu \xi\, U_+(\xi)+\iu q\, V_+(\xi)}{q^2+\xi^2}=\frac{\iu \xi\,C_1 e^{Q_+(\xi)}+\iu q\,C_2 e^{R_+(\xi)}}{q^2+\xi^2}~,\label{eq:u-soln-plus}\\
\widehat v_+(\xi)&=-\frac{\iu q\,U_+(\xi)-\iu \xi\,V_+(\xi)}{q^2+\xi^2}=\frac{\iu q\,C_1 e^{Q_+(\xi)}-\iu \xi\,C_2e^{R_+(\xi)}}{q^2+\xi^2}~, \label{eq:v-soln-plus}
\end{align}
\end{subequations}
in regard to $u_<(x)$ and $v_<(x)$. Notice the appearance of the factor $(\xi^2+q^2)^{-1}$.

Now define 
\begin{equation*}
{\rm sg}(q):=\left\{
\begin{array}{lr}
1 & \ \mbox{if}\ \Re\,q>0~,\cr
-1 & \ \mbox{if}\ \Re\,q<0~,
\end{array}
\right.
\end{equation*}
which is the signum function for $\Re\,q$.
Since $\widehat u_-(\xi)$ and $\widehat v_-(\xi)$ are analytic in $\mathbb{C}_-$, by~\eqref{eq:uv-soln-minus} we impose the conditions that $\iu \xi\,C_1 e^{-Q_-(\xi)}+\iu q\,C_2 e^{-R_-(\xi)}=0$ and $\iu q\,C_1 e^{-Q_-(\xi)}-\iu \xi\,C_2e^{-R_-(\xi)}=0$ at $\xi=-\iu q\,{\rm sg}(q)$, which entail the relation
\begin{subequations}\label{eq:C12-conds}
\begin{equation}
C_1 e^{-Q_-(-\iu q\,{\rm sg}(q))} +\iu\, {\rm sg}(q)\,C_2 e^{-R_-(-\iu q\,{\rm sg}(q))}=0\quad \mbox{if}\ \Re\, q\neq 0~.\label{eq:C12-minus}
\end{equation}
Another condition should be dictated at $\xi=\iu q\, {\rm sg}(q)$ by use of $Q_+$ and $R_+$. By~\eqref{eq:uv-soln-plus} we require that 
$\iu \xi C_1 e^{Q_+(\xi)}+\iu q C_2 e^{R_+(\xi)}$ and $\iu q C_1 e^{Q_+(\xi)}-\iu \xi C_2 e^{R_+(\xi)}$ vanish at $\xi=\iu q\,{\rm sg}(q)$.  
Thus, 
\begin{equation}
C_1e^{Q_+(\iu q\, {\rm sg}(q))}-\iu\, {\rm sg}(q) C_2 e^{R_+(\iu q\, {\rm sg}(q))}=0~, \ \Re\, q\neq 0~. \label{eq:C12-plus}
\end{equation}
\end{subequations}
Equations~\eqref{eq:C12-conds} form a linear system for $(C_1, C_2)$. For nontrivial solutions of this system, we require that
\begin{subequations}\label{eqs:dispersion}
\begin{equation}\label{eq:EP-disp-exp}
e^{R_+(\iu q\,{\rm sg}(q))-Q_-(-\iu q\,{\rm sg}(q))}+e^{Q_+(\iu q\,{\rm sg}(q))-R_-(-\iu q\,{\rm sg}(q))}=0~,
\end{equation}
which is recast to the expression 
\begin{equation}\label{eq:EP-dispersion-I}
\{Q_+(\iu q\,{\rm sg}(q))+Q_-(-\iu q\, {\rm sg}(q))\}-\{R_+(\iu q\,{\rm sg}(q))+R_-(-\iu q\,{\rm sg}(q))\}=\iu (2l+1)\pi
\end{equation}
\end{subequations}
for any $l$ in $\mathbb{Z}$. Equations~\eqref{eqs:dispersion} form our core result. Recall that $Q_\pm(\xi)$ and $R_\pm(\xi)$ are defined by~\eqref{eq:QR-def}. By virtue of~\eqref{eq:C12-conds} and~\eqref{eqs:dispersion}, the constants $C_1$ and $C_2$ are interrelated, as expected.
\medskip

{\em Remark~2. 
Dispersion relation~\eqref{eq:EP-disp-exp} or~\eqref{eq:EP-dispersion-I} exhibits reflection symmetry with respect to $q$, i.e., it is invariant under the replacement $q\to -q$, as anticipated for the case with an isotropic surface conductivity. 
If $\sigma^*(\omega)=-\sigma(\omega)$ and the ambient medium is lossless, we can verify that if $q(\omega)$ is a solution of ~\eqref{eq:EP-dispersion-I} so is $q^*(\omega)$; thus, if $q(\omega)$ is unique for $\Re\,q(\omega)>0$ or $\Re\,q(\omega)<0$, with fixed $\omega$, this $q(\omega)$ must be real.}
\medskip 

The integer $l$ that appears in~\eqref{eq:EP-dispersion-I} deserves some attention. 
\medskip

{\em Remark~3. 
Because $Q_\pm(\xi)$ and $R_\pm(\xi)$ are analytic and single valued, only one value of the integer $l$ is relevant in dispersion relation~\eqref{eq:EP-dispersion-I}; cf. Refs.~\onlinecite{VolkovMikhailov1988,MMSLL-preprint} for a similar discussion. This $l$ should be chosen in conjunction with the branch for the logarithm in the integrals for $Q_\pm$ and $R_\pm$. Of course, relation~\eqref{eq:EP-dispersion-I} should furnish physically anticipated  results. \color{black} For example, $q$ approaches the known quasi-electrostatic limit if $|q|\gg |k_0|$ and $\Im\,\sigma>0$ (Section~\ref{sec:qs} and Ref.~\onlinecite{VolkovMikhailov1988}); also, $q$ should approach $k_0$ at low enough frequencies and thus yield a gapless energy spectrum $\omega(q)$ of the EP in the dissipationless case, if $q$ is real (Section~\ref{sec:small-q} and Ref.~\onlinecite{VolkovMikhailov1988}). We choose to set $l=0$ which implies that the branch of the logarithm $w=\ln\mathcal P_\varpi(\xi)$ ($\varpi={\rm TM}, {\rm TE}$) in the integrals for $Q_\pm$ and $R_\pm$ is such that $-\pi<\Im\,w\le \pi$, when $\xi$ lies in the top Riemann sheet (see Sections~\ref{sec:small-q} and~\ref{sec:qs}).
}
\medskip

{\em Remark 4. Equations~\eqref{eqs:dispersion} express the combined effect of TM and TE polarizations via the terms $Q_\pm(\pm \iu q{\rm sg}(q))$ and $R_\pm(\pm \iu q{\rm sg}(q))$, respectively. In the nonretarded frequency regime, the $R_\pm$ terms become relatively small (see Section~\ref{sec:qs} for details). 

}

\section{Tangential electric field and bulk surface plasmons}
\label{subsec:EP_and_SP}
In this section, we compute the electric field tangential to the
sheet. The EP wave number, $q$, satisfies dispersion relation~\eqref{eqs:dispersion}. 
For definiteness, we henceforth assume that 
\begin{equation*}
\Re\,q>0\ \mbox{and}\ \Im\,q\ge 0~.
\end{equation*}

First, by~\eqref{eq:uv-soln-minus}--\eqref{eq:C12-conds} we obtain the Fourier transforms
\begin{equation*}
\widehat{u}_-(\xi)=-C_1 \left[\iu \xi\, e^{-Q_-(\xi)}-q\, e^{-Q_-(-\iu q)} e^{R_-(-\iu q)-R_-(\xi)}\right] (q^2+\xi^2)^{-1}~,	\end{equation*}
\begin{equation*}
\widehat{v}_-(\xi)=-C_1	\left[ \iu q\, e^{-Q_-(\xi)}+\xi\, e^{-Q_-(-\iu q)} e^{R_-(-\iu q)-R_-(\xi)}\right] (q^2+\xi^2)^{-1}~; 
\end{equation*}
and
\begin{equation*}
\widehat{u}_+(\xi)=C_1  \left[ \iu \xi\,e^{Q_+(\xi)}+q\,e^{Q_+(\iu q)} e^{R_+(\xi)-R_+(\iu q)}\right] (q^2+\xi^2)^{-1}~,
\end{equation*}
\begin{equation*}
	\widehat{v}_+(\xi)= C_1 \left[ \iu q\,e^{Q_+(\xi)}-\xi e^{Q_+(\iu q)} e^{R_+(\xi)-R_+(\iu q)}\right] (q^2+\xi^2)^{-1}~. 
\end{equation*}
These functions are analytic at $\xi=\pm \iu q$.
The inverse Fourier transforms are
\begin{subequations}\label{eqs:Exy-invFT-sheet}
\begin{align}
E_x(x,0)&=-\frac{C_1}{2\pi \iu} \int_{-\infty}^\infty \dxi\ \frac{e^{\iu \xi x}}{q^2+\xi^2} \left[\iu \xi \, e^{-Q_-(\xi)}-q\, e^{-Q_-(-\iu q)} e^{R_-(-\iu q)-R_-(\xi)}\right]~,  \label{eq:Ex-invFT-sheet}	\\
E_y(x,0)&=-\frac{C_1}{2\pi \iu} \int_{-\infty}^\infty \dxi\ \frac{e^{\iu \xi x}}{q^2+\xi^2} \left[\iu q \, e^{-Q_-(\xi)}+\xi\, e^{-Q_-(-\iu q)} 
e^{R_-(-\iu q)-R_-(\xi)}\right]\quad x>0~;  \label{eq:Ey-invFT-sheet}	\end{align}	
\end{subequations}
and
\begin{subequations}\label{eqs:Exy-invFT-outsheet}
\begin{align}
E_x(x,0)&=\frac{C_1}{2\pi \iu} \int_{-\infty}^\infty \dxi\ 
\frac{e^{\iu \xi x}}{q^2+\xi^2} \left[\iu \xi \, e^{Q_+(\xi)}+q\, 
e^{Q_+(\iu q)} e^{R_+(\xi)-R_+(\iu q)}\right]~,  \label{eq:Ex-invFT-outsheet}	\\
E_y(x,0)&=\frac{C_1}{2\pi \iu} \int_{-\infty}^\infty \dxi\ 
\frac{e^{\iu \xi x}}{q^2+\xi^2} \left[\iu q \, e^{Q_+(\xi)}-\xi\, 
e^{Q_+(\iu q)} 
e^{R_+(\xi)-R_+(\iu q)}\right]\quad x<0~.  \label{eq:Ey-invFT-outsheet}	\end{align}	
\end{subequations}
The task now is to approximately evaluate the above integrals for fixed $q$ in the following regimes: (i) $|qx|\ll 1$, close to the edge (Section~\ref{subsec:near-f}); and (ii) for sufficiently large $|qx|$ if $x>0$ (Section~\ref{subsec:far-f}). We describe two types of plausibly emerging SPs, which for fixed $q$ and $\omega$ have distinct propagation constants in the $x$-direction, on the sheet away from the edge. For large $|qx|$, our calculation indicates the localization of the EP on the sheet near the material edge. \color{black}

\subsection{Tangential electric field near the edge, $|qx|\ll 1$}
\label{subsec:near-f}
Consider $x>0$, for points on the sheet. In~\eqref{eqs:Exy-invFT-sheet}, we shift the integration path in the lower half $\xi$-plane, keeping in mind that the integrands are analytic at $\xi=-\iu q$, and write  
\begin{align*}
E_x(x,0)&=-\frac{C_1}{2\pi \iu} \int_{-\infty-\iu \delta_1}^{+\infty-\iu\delta_1} \dxi\ \frac{e^{\iu \xi x}}{q^2+\xi^2} \left[\iu \xi \, e^{-Q_-(\xi)}-q\, e^{-Q_-(-\iu q)} e^{R_-(-\iu q)-R_-(\xi)}\right]~,  \\
E_y(x,0)&=-\frac{C_1}{2\pi \iu} \int_{-\infty-\iu\delta_1}^{+\infty-\iu\delta_1} \dxi\ \frac{e^{\iu \xi x}}{q^2+\xi^2} \left[\iu q \, e^{-Q_-(\xi)}+\xi\, e^{-Q_-(-\iu q)} 
e^{R_-(-\iu q)-R_-(\xi)}\right]\quad x>0~,
\end{align*}	
for a positive constant $\delta_1$ with $\delta_1\gg |q|$ and $\delta_1 x \ll 1$. Thus, the factor $e^{\iu \xi x}$ in each integrand has a magnitude close to unity.
From the Appendix, we use the asymptotic formulas 
\begin{equation*}
e^{-Q_-(\xi)}=\mathfrak Q(\xi)[1+o(1)]\quad \mbox{and}\quad  
e^{-R_-(\xi)}=1+o(1)\ \mbox{as}\ \xi\to\infty~,
\end{equation*}
where
\begin{equation*}
\mathfrak Q(\xi):= \left(-\frac{\omega\mu\sigma\xi}{2k_0^2}\right)^{-1/2}
\end{equation*}
which has a branch cut emanating from the origin in $\mathbb{C}_+$.

The component of the electric field parallel to the edge on the sheet approaches the limit 
\begin{align*}
\lim_{x\downarrow 0}E_y(x,0)&=  -\frac{C_1}{2\pi\iu}\left[\iu q\int_{-\infty-\iu\delta_1}^{+\infty-\iu\delta_1}\dxi\,(q^2+\xi^2)^{-1}\,e^{-Q_-(\xi)} \right. \notag \\
& \quad \left. +e^{-Q_-(-\iu q)+R_-(-\iu q)}\lim_{x\downarrow 0}\int_{-\infty-\iu\delta_1}^{+\infty-\iu\delta_1}\dxi\,
e^{\iu \xi x} (q^2+\xi^2)^{-1}\xi e^{-R_-(\xi)}	\right]~.
\end{align*}
Since $e^{-Q_-(\xi)}=\mathcal O(\xi^{-1/2})$ as $\xi\to\infty$, we infer that the first one of the above integrals converges. In fact, we see that this integral vanishes by closing the integration path through a large semicircle in $\mathbb{C}_-$. The second integral is evaluated via the approximations 
$q^2+\xi^2\sim \xi^2$ and $e^{-R_-(\xi)}\sim 1$ since $|\xi|\gg |q|$.
Hence, at the edge $E_y(x,0)$ on the sheet has the finite value
\begin{equation}\label{eq:Ey-edge-sheet}
\lim_{x\downarrow 0}E_y(x,0)=:E_y(0^+,0)=-C_1 e^{-Q_-(-\iu q)+R_-(-\iu q)}~,	\end{equation}
where $q$ satisfies~\eqref{eqs:dispersion}; cf. Ref.~\onlinecite{VolkovMikhailov1988} in the context of the quasi-electrostatic approach. It can be shown that the correction to this leading-order term for $E_y(x,0)$ is of the order of $|k_0 x|$.

In a similar vein, we can address $E_x(x,0)$, the component of the electric field on the sheet normal to the edge.
Without further ado, we compute (with $\delta_1\gg |q|$)
\begin{align*}
E_x(x,0)&\sim -\frac{C_1}{2\pi\iu}\left[\int_{-\infty-\iu\delta_1}^{+\infty-\iu\delta_1}\dxi\ \frac{e^{i\xi x}}{q^2+\xi^2}\,(\iu\xi)\,\mathfrak Q(\xi)-q e^{-Q_-(-\iu q)+R_-(-\iu q)}\int_{-\infty-\iu\delta_1}^{+\infty-\iu\delta_1}\dxi\ \frac{e^{-R_-(\xi)}}{q^2+\xi^2}\right]\\
&= -\frac{C_1}{2\pi\iu} \int_{-\infty-\iu\delta_1}^{+\infty-\iu\delta_1}\dxi\ \frac{e^{i\xi x}}{q^2+\xi^2}\,(\iu\xi)\,\mathfrak Q(\xi)\sim -\frac{C_1}{2\pi}\int_{-\infty-\iu\delta_1}^{+\infty-\iu\delta_1}\dxi\ \frac{e^{i\xi x}}{\xi}\,\mathfrak Q(\xi)~.
\end{align*}
By applying integration by parts once and wrapping the integration contour around the positive imaginary axis in the $\xi$-plane, we obtain 
\begin{equation}\label{eq:Ex-edge-sheet}
E_x(x,0)\sim 2C_1 \left(\frac{2\iu\,k_0}{\pi\omega\mu\sigma}\right)^{1/2}\sqrt{k_0 x}~,\quad |qx|\ll 1~,\ x>0~.
\end{equation} 
Thus, the surface current normal to the edge vanishes, as it happens also for line currents at the ends of cylindrical antennas with a delta-function voltage generator.~\cite{KingFikioris2002} For a similar result in the scattering of waves from conducting films, see equation (39) in Ref.~\onlinecite{MML2017}.

Consider $x<0$, if the observation point lies at $z=0$ outside the sheet. By~\eqref{eqs:Exy-invFT-outsheet}, we have
\begin{align*}
E_x(x,0)&=\frac{C_1}{2\pi \iu} \int_{-\infty+\iu\delta_1}^{+\infty+\iu\delta_1} \dxi\ 
\frac{e^{\iu \xi x}}{q^2+\xi^2} \left[\iu \xi \, e^{Q_+(\xi)}+q\, 
e^{Q_+(\iu q)} e^{R_+(\xi)-R_+(\iu q)}\right]~,  \\
E_y(x,0)&=\frac{C_1}{2\pi \iu} \int_{-\infty+\iu\delta_1}^{+\infty+\iu\delta_1} \dxi\ 
\frac{e^{\iu \xi x}}{q^2+\xi^2} \left[\iu q \, e^{Q_+(\xi)}-\xi\, 
e^{Q_+(\iu q)} 
e^{R_+(\xi)-R_+(\iu q)}\right]\quad x<0~, 
\end{align*}	
where $\delta_1\gg |q|$ and $\delta_1 |x|\ll 1$. We will also need the following formulas (see Appendix):
\begin{equation*}
e^{Q_+(\xi)}=\mathfrak Q(-\xi)^{-1}[1+o(1)]\quad \mbox{and}\quad e^{R_+(\xi)}=1+o(1)\ \mbox{as}\ \xi\to\infty~,
\end{equation*}
noting that $\mathfrak Q(-\xi)$ has a branch cut emanating from the origin in $\mathbb{C}_-$.

For $|qx|\ll 1$, we therefore compute 
\begin{align}\label{eq:Ey-edge-out}
\lim_{x\uparrow 0}E_y(x,0)&=:E_y(0-,0)= \frac{C_1}{2\pi\iu}\left[\iu q\int_{-\infty+\iu\delta_1}^{+\infty+\iu\delta_1}\dxi\ \frac{e^{Q_+(\xi)}}{q^2+\xi^2}\right. \notag \\
& \quad \left. -e^{Q_+(\iu q)-R_+(\iu q)}\lim_{x\uparrow 0}\int_{-\infty+\iu\delta_1}^{+\infty+\iu\delta_1}\dxi\ e^{i\xi x}\,\frac{\xi}{q^2+\xi^2}\,e^{R_+(\xi)}\right]\notag \\
&= 	-\frac{C_1}{2\pi \iu}\, e^{Q_+(\iu q)-R_+(\iu q)}\lim_{x\uparrow 0}\int_{-\infty+\iu\delta_1}^{+\infty+\iu\delta_1}\dxi\ e^{i\xi x}\,\frac{\xi}{q^2+\xi^2}\,e^{R_+(\xi)}\notag \\
&= C_1\, e^{Q_+(\iu q)-R_+(\iu q)}~.
\end{align}
In the above, the integral of the first line is convergent; in fact, this integral vanishes. In the integrand of the remaining integral, we use the approximations $q^2+\xi^2\sim \xi^2$ and $e^{R_+(\xi)}\sim 1$. 
By dispersion relation~\eqref{eq:EP-disp-exp} and limit~\eqref{eq:Ey-edge-sheet}, we conclude that $E_y(x,0)$ is continuous across the edge, viz.,
\begin{equation*}
E_y(0^-,0)=E_y(0^+,0)~.	
\end{equation*}

On the other hand, the $x$-component of the electric field at $z=0$ outside the sheet is
\begin{align}\label{eq:Ex-edge-out}
E_x(x,0)&\sim\frac{C_1}{2\pi \iu} \left[ \int_{-\infty+\iu\delta_1}^{+\infty+\iu\delta_1} \dxi\ 
\frac{e^{\iu \xi x}}{q^2+\xi^2}\,(\iu \xi) \, \mathfrak Q(-\xi)^{-1} +q\, e^{Q_+(\iu q)-R_+(\iu q)}\int_{-\infty+\iu\delta_1}^{+\infty+\iu\delta_1}\dxi\ \frac{e^{R_+(\xi)}}{q^2+\xi^2}\right] \notag \\
 &= \frac{C_1}{2\pi \iu} \int_{-\infty+\iu\delta_1}^{+\infty+\iu\delta_1} \dxi\ 
\frac{e^{\iu \xi x}}{q^2+\xi^2}\,(\iu \xi) \, \mathfrak Q(-\xi)^{-1} \sim  \frac{C_1}{2\pi} \int_{-\infty+\iu\delta_1}^{+\infty+\iu\delta_1} \dxi\ 
\frac{e^{\iu \xi x}}{\xi}\, \mathfrak Q(-\xi)^{-1}	\notag \\
&= C_1 \left(\frac{\omega\mu\sigma}{2\pi\iu\,k_0}\right)^{1/2} \frac{1}{\sqrt{k_0 |x|}}~,\quad |qx|\ll 1,\ x<0~.
\end{align}
Thus, $\partial E_x(x,z)/\partial x$ indeed has a non-integrable singularity as $x\uparrow 0$ at $z=0$ (see Remark~1).

\subsection{Far field: Two types of bulk SPs in the direction normal to the edge}
\label{subsec:far-f}
Next, we describe the bulk SPs in the $x$-direction with recourse to the Fourier integrals for $E_x(x,0)$ and $E_y(x,0)$. By~\eqref{eq:QR-def} and~\eqref{eqs:Exy-invFT-sheet}, for $x>0$ we use the integral representations
\begin{subequations}\label{eq:Exy-xpos-poles}
\begin{align}
E_x(x,0)&=-\frac{C_1}{2\pi\iu}\int_{-\infty}^\infty \dxi\,\frac{e^{\iu\xi x}}{q^2+\xi^2}\left\{\frac{\iu\xi}{\mathcal P_{\rm TM}(\xi)} e^{Q_+(\xi)}-e^{R_-(-\iu q)-Q_-(-\iu q)}\frac{q}{\mathcal P_{\rm TE}(\xi)} e^{R_+(\xi)}\right\}~,
\end{align}
\begin{align}
E_y(x,0)&=-\frac{C_1}{2\pi\iu}\int_{-\infty}^\infty \dxi\,\frac{e^{\iu\xi x}}{q^2+\xi^2}\left\{\frac{\iu q}{\mathcal P_{\rm TM}(\xi)} e^{Q_+(\xi)}+e^{R_-(-\iu q)-Q_-(-\iu q)}\frac{\xi}{\mathcal P_{\rm TE}(\xi)} e^{R_+(\xi)}\right\}~,
\end{align}
\end{subequations}
where $q$ solves~\eqref{eqs:dispersion}, and $\mathcal P_{\rm TM}(\xi)$ and $\mathcal P_{\rm TE}(\xi)$ are defined by~\eqref{eq:K-eigenv}. Note that the integrands are analytic at $\xi=\iu q$. 
\medskip

{\em Definition~2 (2D bulk SPs). Consider the electric field tangential to the sheet. For every $q$ solving~\eqref{eqs:dispersion} with given $\omega$, the {\rm 2D bulk SPs} in the positive $x$-direction 
are identified with waves that arise from~\eqref{eq:Exy-xpos-poles} as residues of the integrands from the zeros of $\mathcal P_{\rm TM}(\xi)$ or $\mathcal P_{\rm TE}(\xi)$ in the upper half $\xi$-plane of the top Riemann sheet ($\Re\sqrt{\xi^2+q^2-k_0^2}>0$).

If the ambient medium is lossless ($k_0>0$) we can characterize these waves as follows. If $\Im\,\sigma(\omega)>0$, only the zeros $\xi=\pm k_{\rm sp}^{\rm e}$ ($k_{\rm sp}^{\rm e}\in\mathbb{C}_+$) of $\mathcal P_{\rm TM}(\xi)$ are present in the top Riemann sheet; see~\eqref{eq:ksp-TM} below. In this case, only $k_{\rm sp}^{\rm e}$ contributes to the residues, which amounts to a TM-like bulk SP.~\cite{Bludov2013,ML2016,MML2017} Similarly, if $\Im\,\sigma(\omega)<0$ only the zero $\xi=k_{\rm sp}^{\rm m}\in\mathbb{C}_+$ of $\mathcal P_{\rm TE}(\xi)$ contributes to the residues; see~\eqref{eq:ksp-TE} below. This case signifies a TE-like bulk SP.~\cite{Bludov2013,ML2016,MML2017} 
}
\medskip

Definition~2 does not explain how these 2D SPs can be separated from other contributions to the Fourier integrals for $E_x(x,0)$ and $E_y(x,0)$. We address this issue in a simplified way.

By~\eqref{eq:Exy-xpos-poles} we proceed to calculate $E_x(x,0)$ and $E_y(x,0)$ by contour integration in the far field, for sufficiently large $|\sqrt{q^2-k_0^2}\,x|$, and thus indicate the emergence of bulk SPs as possibly distinct contributions. By closing the path in the upper half $\xi$-plane, we write
\begin{equation*}
E_\ell(x,0)=E_\ell^{\rm sp}(x,0)+E_\ell^{\rm rad}(x,0)\qquad (\ell=x,\,y)~,
\end{equation*}
where $E_\ell^{\rm sp}$ is the residue contribution, which amounts to a bulk SP in the $x$-direction (Definition~2), and $E_\ell^{\rm rad}$ is the contribution from the branch cut emanating from the point $\iu \sqrt{q^2-k_0^2}$ ($\Re\sqrt{q^2-k_0^2}>0$). We refer to the latter contribution as the `radiation field'.~\cite{MML2017} In this simplified treatment, we focus on large enough distances from the edge so that the relevant pole contribution is sufficiently separated from the branch point contribution.

First, we consider $E_\ell^{\rm sp}(x,0)$ ($\ell=x,\,y$). After some algebra, for $k_0>0$ we obtain 
\begin{subequations}\label{eqs:residues-bulk-SPs}
\begin{equation}
\frac{E_x^{\rm sp}(x,0)}{C_1}=
\left\{ \begin{array}{lr}
{\displaystyle \iu\biggl[1-\left(\frac{\omega\mu\sigma}{2k_0}\right)^2\biggr]^{-1} e^{Q_+(k_{\rm sp}^{\rm e})} e^{ik_{\rm sp}^{\rm e}x}}~,& \Im\sigma>0~,\\
{\displaystyle \frac{q}{k_{\rm sp}^{\rm m}}\biggl(\frac{\omega\mu\sigma}{2k_0}\biggr)^2\biggl[1-\left(\frac{\omega\mu\sigma}{2k_0}\right)^2\biggr]^{-1} e^{R_+(k_{\rm sp}^{\rm m})-R_+(\iu q)+Q_+(\iu q)} e^{ik_{\rm sp}^{\rm m}x}}~, &  \Im\sigma<0~;
\end{array} \right.
\end{equation}
\begin{equation}
\frac{E_y^{\rm sp}(x,0)}{C_1}= 
\left\{ \begin{array}{lr}
{\displaystyle \frac{\iu q}{k_{\rm sp}^{\rm e}}\biggl[1-\left(\frac{\omega\mu\sigma}{2k_0}\right)^2\biggr]^{-1} e^{Q_+(k_{\rm sp}^{\rm e})} e^{ik_{\rm sp}^{\rm e}x}}~,& \Im\sigma>0~, \\
{\displaystyle -\biggl(\frac{\omega\mu\sigma}{2k_0}\biggr)^2\biggl[1-\left(\frac{\omega\mu\sigma}{2k_0}\right)^2\biggr]^{-1} e^{R_+(k_{\rm sp}^{\rm m})-R_+(\iu q)+Q_+(\iu q)} e^{ik_{\rm sp}^{\rm m}x}}~, & \Im\sigma<0~.
\end{array} \right.
\end{equation}
\end{subequations}
In the above, from the zeros of $\mathcal P_{\rm TM}(\xi)$ and $\mathcal P_{\rm TE}(\xi)$ we define the wave numbers 
\begin{subequations}\label{eq:ksp-TM-TE}
\begin{align}
k_{\rm sp}^{\rm e}&=\iu\sqrt{q^2-\biggl(\frac{\iu 2k_0^2}{\omega\mu\sigma}\biggr)^2-k_0^2}\quad \mbox{if}\ \Im\,\sigma>0\ 
(\mbox{TM})~,\label{eq:ksp-TM} \\
k_{\rm sp}^{\rm m}&=\iu \sqrt{q^2-\biggl(\frac{\omega\mu\sigma}{2\iu}\biggr)^2-k_0^2}\quad (\Im\,k_{\rm sp}^{\rm e, m}>0)\quad \mbox{if}\ \Im\,\sigma<0\ (\mbox{TE})~,\label{eq:ksp-TE}
\end{align}
\end{subequations}
so that $k_{\rm sp}^{\rm e}$ or $k_{\rm sp}^{\rm m}$ lies in the top Riemann sheet, respectively.~\cite{ML2016,MML2017}  
\medskip

{\em Remark~5. 
In the nonretarded frequency regime (Section~\ref{sec:qs}),
if $|\omega\mu\sigma/k_0|\ll 1$ and $\Im\,\sigma(\omega)>0$, the $q$ that solves~\eqref{eqs:dispersion} for fixed $\omega$ is given by $q\sim \eta_0\, (\iu 2k_0^2/(\omega\mu\sigma))$
with $\eta_0>1$;\cite{VolkovMikhailov1985,VolkovMikhailov1988} thus, the TM-like SP is significantly damped. In this regime,  we approximate $k_{\rm sp}^{\rm e}\sim \iu\sqrt{q^2-[\iu 2k_0^2/(\omega\mu\sigma)]^2}$. Note that this approximation can be used in the exponential factor for $E_x^{\rm sp}(x,0)$ and $E_y^{\rm sp}(x,0)$ with a small error if $|\omega\mu\sigma|x\ll 1$, along with $|\omega\mu\sigma/k_0|\ll 1$. }
\medskip

Next, we calculate the contributions, $E_\ell^{\rm rad}(x,0)$, along the branch cut ($\ell=x,\,y$). Suppose that $\sqrt{q^2-k_0^2}>0$. By the change of variable $\xi\mapsto \varsigma$ with $\xi=\iu \sqrt{q^2-k_0^2}(1+\varsigma)$ and $\varsigma>0$, we express the requisite integrals as
\begin{align*}
E_x^{\rm rad}(x,0)&= -\frac{C_1}{2\pi}\biggl\{\frac{\omega\mu\sigma}{k_0}\frac{\sqrt{q^2-k_0^2}}{k_0}\int_0^\infty {\rm d}\varsigma\ \frac{e^{-\sqrt{q^2-k_0^2}\,x\varsigma} e^{Q_+(\iu\sqrt{q^2-k_0^2}(1+\varsigma))}}{(1+\varsigma)^2-\frac{q^2}{q^2-k_0^2}}
\frac{(1+\varsigma)\sqrt{\varsigma (2+\varsigma)}}{1-\bigl(\frac{\omega\mu\sigma}{2k_0}\bigr)^2\frac{q^2-k_0^2}{k_0^2}\varsigma (2+\varsigma)}\\
&\mbox{} +4e^{-R_+(\iu q)+Q_+(\iu q)}\frac{q}{\omega\mu\sigma}\int_0^\infty {\rm d}\varsigma\ 
\frac{e^{-\sqrt{q^2-k_0^2}\,x\varsigma}}{(1+\varsigma)^2-\frac{q^2}{q^2-k_0^2}}
\frac{e^{R_+(\iu\sqrt{q^2-k_0^2}(1+\varsigma))}}{1-\bigl(\frac{2}{\omega\mu\sigma}\bigr)^2(q^2-k_0^2)\varsigma(2+\varsigma)} \sqrt{\varsigma(2+\varsigma)}\biggr\}\\
&\mbox{}\qquad \times e^{-\sqrt{q^2-k_0^2}\,x}
\end{align*} 
and
\begin{align*}
E_y^{\rm rad}(x,0)&= \frac{\iu C_1}{2\pi}\biggl\{\frac{q}{k_0}\frac{\omega\mu\sigma}{k_0}\int_0^\infty {\rm d}\varsigma\ \frac{e^{-\sqrt{q^2-k_0^2}\,x\varsigma} e^{Q_+(\iu\sqrt{q^2-k_0^2}(1+\varsigma))}}{(1+\varsigma)^2-\frac{q^2}{q^2-k_0^2}}
\frac{\sqrt{\varsigma(2+\varsigma)}}{1-\bigl(\frac{\omega\mu\sigma}{2k_0}\bigr)^2\frac{q^2-k_0^2}{k_0^2}\varsigma (2+\varsigma)}\\
&\mbox{} +4e^{-R_+(\iu q)+Q_+(\iu q)}\frac{\sqrt{q^2-k_0^2}}{\omega\mu\sigma}\int_0^\infty {\rm d}\varsigma\ 
\frac{e^{-\sqrt{q^2-k_0^2}\,x\varsigma}}{(1+\varsigma)^2-\frac{q^2}{q^2-k_0^2}}
\frac{e^{R_+(\iu\sqrt{q^2-k_0^2}(1+\varsigma))} (1+\varsigma) \sqrt{\varsigma (2+\varsigma )}}{1-\bigl(\frac{2}{\omega\mu\sigma}\bigr)^2(q^2-k_0^2)\varsigma (2+\varsigma)}\biggr\}\\
&\mbox{}\qquad \times e^{-\sqrt{q^2-k_0^2}\,x}~.
\end{align*} 
In the far field, when $|\sqrt{q^2-k_0^2}\,x|\gg 1$ with 
\begin{equation*}
\biggl|\biggl(\frac{\omega\mu\sigma}{2k_0}\biggr)^2\frac{q^2-k_0^2}{k_0^2}\frac{1}{\sqrt{q^2-k_0^2}x}\biggr|\ll 1\quad \mbox{and}\quad \biggl|\frac{q^2-k_0^2}{(\omega\mu\sigma)^2}\frac{1}{\sqrt{q^2-k_0^2}\,x}\biggr|\ll 1,
\end{equation*}
the major contribution to integration in the above branch cut integrals comes from the endpoint, $\varsigma=0$. Accordingly, we evaluate 
\begin{subequations}\label{eqs:rad-field-approx}
\begin{align}
E_x^{\rm rad}(x,0)&\sim \frac{C_1}{\sqrt{2\pi}}\frac{q^2-k_0^2}{k_0^2}
\biggl\{\frac{\omega\mu\sigma}{2k_0}\frac{\sqrt{q^2-k_0^2}}{k_0} e^{Q_+(\iu\sqrt{q^2-k_0^2})}+2\frac{q}{\omega\mu\sigma}
e^{-R_+(\iu q)+Q_+(\iu q)}e^{R_+(\iu\sqrt{q^2-k_0^2})}\biggr\}\notag\\
&\mbox{}\quad \times \frac{e^{-\sqrt{q^2-k_0^2}\,x}}{(\sqrt{q^2-k_0^2} x)^{3/2}}~,
\end{align}
\begin{align}
E_y^{\rm rad}(x,0)&\sim -\frac{\iu C_1}{\sqrt{2\pi}}\frac{q^2-k_0^2}{k_0^2}
\biggl\{\frac{\omega\mu\sigma}{2k_0}\frac{q}{k_0} e^{Q_+(\iu\sqrt{q^2-k_0^2})}+2\frac{\sqrt{q^2-k_0^2}}{\omega\mu\sigma}
e^{-R_+(\iu q)+Q_+(\iu q)}e^{R_+(\iu\sqrt{q^2-k_0^2})}\biggr\}\notag\\
&\mbox{}\quad \times \frac{e^{-\sqrt{q^2-k_0^2}\,x}}{(\sqrt{q^2-k_0^2} x)^{3/2}}\qquad (x>0)~.
\end{align}
\end{subequations}
The above far-field formulas for $E_x^{\rm rad}(x,0)$ and $E_y^{\rm rad}(x,0)$ can be analytically continued to complex $\sqrt{q^2-k_0^2}$ with $\Re\sqrt{q^2-k_0^2}>0$. 
\medskip

{\em Remark~6. By the formulas for $E_\ell^{\rm rad}(x,0)$ ($\ell=x, y$), this contribution may decay rapidly with $x$ if $|q|\gg k_0$. This can occur in the nonretarded frequency regime (see Section~\ref{sec:qs}), where $\Im\,\sigma>0$ and $q\simeq \eta_0 (\iu 2k_0^2/(\omega\mu\sigma))$ with $\eta_0>1$.~\cite{VolkovMikhailov1988}
By inspection of the simplified formulas for the TM-like bulk SP, $E_\ell^{\rm sp}$, and the radiation field $E_\ell^{\rm rad}$, we expect that, in the nonretarded frequency regime, the SP contribution can be dominant over the radiation field. Hence, the EP electric field tangential to the sheet can be localized near the edge on the 2D material.}
\medskip

A more accurate study of the electric field would involve the derivation of asymptotic formulas for the requisite Fourier integrals in an intermediate regime of distances from the edge, between the near and far fields. In addition, the $q(\omega)$ must be numerically computed from dispersion relation~\eqref{eqs:dispersion} for various material parameters and frequencies of interest. These tasks will be the subject of future work.\cite{MMSLL-inprep}

\section{On the low-frequency EP dispersion relation}
\label{sec:small-q}
In this section, we derive an asymptotic formula for the $q$ that obeys~\eqref{eqs:dispersion} if
\begin{subequations}\label{eq:low-freq-conds}
\begin{equation}\label{eq:low-freq-condI}
\biggl|\frac{\omega\mu\sigma(\omega)}{k_0}\biggr|\gg 1\quad \mbox{and}\quad \Im\,\sigma(\omega)> 0\qquad (k_0=\omega\sqrt{\mu\varepsilon})~.
\end{equation}
One way to motivate these conditions is to invoke the Drude model for doped single-layer graphene, which is expected to be accurate for small enough plasmon energies.\cite{Jablan2013} By this model, $\sigma(\omega)=\iu [e^2 v_F \sqrt{n_s}/(\sqrt{\pi}\hbar)](\omega+\iu/\tau_e)^{-1}$; $e$ is the electron charge, $v_F$ is the Fermi velocity, $\tau_e$ is the relaxation time of microscopic collisions, $n_s$ is the electron surface density, and $\hbar$ is Planck's constant, while the interband transitions are neglected in the calculation of this $\sigma(\omega)$.\cite{Falkovsky2007} Hence, within this model, the conditions of~\eqref{eq:low-freq-condI} are obeyed if
\begin{equation}\label{eq:low-freq-condII}
\tau_e^{-1}\ll \omega\ll \omega_p\quad \mbox{where}\ \omega_p=\biggl|Z_0\frac{e^2\sqrt{n_s}v_F}{\sqrt{\pi}\hbar}\biggl|~,\quad Z_0= \sqrt{\frac{\mu}{\varepsilon}}~;
\end{equation}
\end{subequations}
$Z_0$ is the characteristic impedance of the (unbounded) ambient medium. For given $\omega$, the conditions in~\eqref{eq:low-freq-condII} call for large enough relaxation time, $\tau_e$, and surface density, $n_s$. We expect that $q/k_0=\mathcal O(1)$ with $|q|> k_0$ in this regime. Our task here is to refine this anticipated result. Note that the model for $\sigma(\omega)$ can be improved by consideration of both the intraband and interband transitions in the linear-response quantum theory for $\sigma$.\cite{Falkovsky2007}

First, we convert~\eqref{eq:EP-dispersion-I} for the EP dispersion to a more explicit expression with $\Re\,q>0$. Consider integral formulas~\eqref{eq:QR-def} for $Q_\pm(\xi)$. By changing the integration variable, $\xi'$, according to $\xi'=q\,\varsigma$, we can alternatively write~\eqref{eq:EP-dispersion-I}, with $l=0$, as 
\begin{align}\label{eq:EP-dispersion-alt}
I(q):=&\frac{2}{\pi}\int_0^\infty \frac{{\rm d}\varsigma}{1+\varsigma^2}\ \left\{
\ln\left[1+\frac{\iu \omega\mu\sigma}{2k_0}\frac{q}{k_0}
\left(\varsigma^2+1-k_0^2/q^2\right)^{1/2}\right] \right. \notag \\
& \qquad \left. -\ln\left[1-\frac{\iu \omega\mu\sigma}{2q}
\left(\varsigma^2+1-k_0^2/q^2\right)^{-1/2}\right]\right\}= \iu\pi~;
\end{align}
\begin{equation*}
\Re\sqrt{q^2(\varsigma^2+1)-k_0^2}>0~.
\end{equation*}
The last condition defines the top Riemann sheet in the $\varsigma$-plane for the integrand in~\eqref{eq:EP-dispersion-alt}.
By use of~\eqref{eq:low-freq-condI}, we notice that in the present frequency regime we have  
\begin{align*}
I(q)&=\frac{2}{\pi}\int_0^\infty \frac{{\rm d}\varsigma}{1+\varsigma^2}\,\biggl\{\ln\biggl(e^{\iu \pi}\biggl(-\frac{\iu\omega\mu\sigma}{2k_0}\biggr)\biggr)-\ln\bigg(-\frac{\iu\omega\mu\sigma}{2k_0}\biggr)\biggr\}+\frac{2}{\pi}\int_0^\infty \frac{{\rm d}\varsigma}{1+\varsigma^2}\, \ln\biggl(\frac{\varsigma^2+\bar q^2}{1-\bar q^2}\biggr)\\
&\qquad +\mathcal O(\epsilon\ln\epsilon)\\
&=\iu\pi +\frac{2}{\pi}\int_0^\infty \frac{{\rm d}\varsigma}{1+\varsigma^2}\, \ln\biggl(\frac{\varsigma^2+\bar q^2}{1-\bar q^2}\biggr) +\mathcal O(\epsilon\ln\epsilon)~;
\quad \epsilon=\iu \frac{2k_0}{\omega\mu\sigma}~,\quad \bar q^2=1-\frac{k_0^2}{q^2}\quad (|\bar q|<1)
\end{align*}
with $|\epsilon|\ll 1$. In the above, we defined the branch of the logarithm, $w=\ln(\cdot)$, by $-\pi<\Im\,w\le \pi$; accordingly, $\bar q\to 0$ as $\epsilon\to 0$, when $q$ approaches $k_0$ (see Remark~3). More generally, we may define  $(2l_0-1)\pi<\Im\,w\le (2l_0+1)\pi$ for some $l_0\in\mathbb{Z}$ while we set $l=l_0$ in~\eqref{eq:EP-dispersion-I}.

We proceed to describe in some detail the asymptotics for~\eqref{eq:EP-dispersion-alt}. By invoking the identity
\begin{align*}
& \ln\biggl(1-\epsilon^{-1}\sqrt{\frac{\varsigma^2+\bar q^2}{1-\bar q^2}}\biggr)-\ln\biggl(1+\epsilon^{-1}\sqrt{\frac{1-\bar q^2}{\varsigma^2+\bar q^2}}\biggr)=\iu\pi+2\ln\biggl(\sqrt{\frac{\varsigma^2+\bar q^2}{1-\bar q^2}}\biggr)\\
&\mbox{}\quad +\ln\biggl(1-\epsilon \sqrt{\frac{1-\bar q^2}{\varsigma^2+\bar q^2}}\biggr)-\ln\biggl(1 +\epsilon \frac{\varsigma}{\sqrt{1-\bar q^2}} +\epsilon\frac{\sqrt{\varsigma^2+\bar q^2}-\varsigma}{\sqrt{1-\bar q^2}}\biggr)~,
\end{align*}
and treating the term $\epsilon(\sqrt{\varsigma^2+\bar q^2}-\varsigma)/\sqrt{1-\bar q^2}$ as a perturbation in the last logarithm, we approximate~\eqref{eq:EP-dispersion-alt} by the relation
\begin{equation*}
	I_1(\bar q)-\bar\epsilon I_2(\bar q)-I_3(\bar\epsilon)\sim  \frac{\pi}{2}\ln(1-\bar q^2)~,\quad \bar\epsilon=\frac{\epsilon}{\sqrt{1-\bar q^2}}~,
\end{equation*}
where
\begin{align*}
	I_1(\chi)&=\int_{0}^\infty {\rm d}\varsigma\ \frac{\ln(\varsigma^2+\chi^2)}{1+\varsigma^2}~,\quad I_2(\chi)=\int_0^\infty {\rm d}\varsigma\,\biggl(\frac{1}{\sqrt{\varsigma^2+\chi^2}}-\frac{\varsigma}{1+\varsigma^2}\biggr)~,\\
	I_3(\chi)&=\int_0^\infty {\rm d}\varsigma\ \frac{\ln(1+\chi\varsigma)}{1+\varsigma^2}~.
\end{align*}
The first two integrals can be computed directly. For $I_1(\chi)$, by contour integration we obtain
\begin{equation*}
	I_1(\chi)=\pi \ln(1+\chi)= \pi \left(\chi-\textstyle{\frac{1}{2}}\chi^2\right)+\mathcal O(\chi^3)~,\quad |\chi|\ll 1~.
\end{equation*}
The integral $I_2(\chi)$ is expressed as
\begin{align*}
I_2(\chi)&=\lim_{M\to\infty}\biggl\{\int_0^M\frac{{\rm d}\varsigma}{\sqrt{\varsigma^2+\chi^2}}-\int_0^M{\rm d}\varsigma\,\frac{\varsigma}{1+\varsigma^2}\biggr\}\\
&=\lim_{M\to\infty}\biggl\{\ln\biggl(\frac{M}{\chi}+\sqrt{1+\frac{M^2}{\chi^2}}\biggr)-\textstyle{\frac{1}{2}}\ln(1+M^2)\biggr\}=\ln (2/\chi)~.
\end{align*}

In order to obtain an asymptotic expansion for $I_3(\chi)$ as $\chi\to 0$ we use the Mellin transform technique.\cite{Sasiela1993} The idea is to compute the Mellin transform, $\widetilde I_3(s)$, of $I_3(\chi)$, and then employ its inversion formula; the desired asymptotic expansion for $I_3(\chi)$ comes from the residues at poles of 
${\widetilde I}_3(s)$ in the $s$-plane with $\Re\,s\ge \alpha$, for some suitable real $\alpha$. For $\chi>0$, define
\begin{equation*}
\widetilde I_3(s)=\int_0^\infty {\rm d}\chi \, I_3(\chi)\chi^{-s}=\frac{1}{2}\frac{\Gamma(s)}{(s-1)^2}\Gamma(2-s)\,\Gamma\left(\frac{s}{2}\right)\,\Gamma\left(-\frac{s}{2}+1\right)~,\quad 1<\Re\,s<2=\alpha
\end{equation*}
so that this integral converges, where $\Gamma(\zeta)$ is the gamma function.\cite{Bateman-I} Here, we interchanged the order of integration (in $\chi$ and $\varsigma$) and used a known integral for the beta function, $B(\zeta_1,\zeta_2)=\Gamma(\zeta_1)\Gamma(\zeta_2)/\Gamma(\zeta_1+\zeta_2)$.\cite{Bateman-I} Consider the inversion formula
\begin{equation*}
I_3(\chi)=\frac{1}{2\pi\iu}\int_{c_1-\iu\infty}^{c_1+\iu\infty} {\rm d}s\ \chi^{s-1} \widetilde I_3(s)~,\qquad 1<c_1<2=\alpha~,
\end{equation*}
and shift the integration path to the right, i.e., into the region of the $s$-plane with $\Re\,s\ge \alpha=2$, noticing that $\widetilde I_3(s)$ has poles at the integers $s=n$ (with $n\ge 2$) in this region. By applying the residue theorem at the double pole $s=2$ and the simple pole $s=3$, we find
\begin{equation*}
I_3(\chi)= \chi (1-\ln\chi)+\frac{\pi}{4}\chi^2+\mathcal O(\chi^3\ln\chi)\qquad \mbox{as}\ \chi\to 0~.
\end{equation*}
This expansion can be analytically continued to complex $\chi$ with $\Re\,\chi\ge 0$.

Consequently, after some algebra, dispersion relation~\eqref{eq:EP-dispersion-alt} is reduced to the formula
\begin{subequations}\label{eqs:q-low-freq}
\begin{equation}\label{eq:q-low-freq-expr}
q\sim k_0\biggl\{1+\frac{1}{2\pi^2}
\epsilon^2\,
\mathcal A(\epsilon)^2\biggr\}~,\quad \epsilon=\frac{\iu\,2k_0}{\omega\mu\sigma}~,
\end{equation}
where for simplicity we neglected terms $o(\epsilon^2)$ on the right-hand side.
In the above, the function $\mathcal A(\epsilon)$ amounts to logarithmic corrections and solves the equation
\begin{equation}\label{eq:q-low-freq-A}
	e^{\mathcal A}=\frac{2e\pi}{\epsilon^2 \mathcal A}~.
\end{equation}
\end{subequations}
Note that an expansion for $\mathcal A(\epsilon)$ can be formally constructed via the iterative scheme
\begin{equation*}
	\mathcal A^{(n+1)}(\epsilon)=\ln\biggl(\frac{2e\pi}{\epsilon^2}\biggr)-\ln\mathcal A^{(n)}(\epsilon)~,\quad \mathcal A^{(0)}(\epsilon)=\ln\frac{2 e\pi}{\epsilon^2}\quad (n=0,\,1,\,\ldots)~.
\end{equation*}
Finally, one can verify that the result furnished by~\eqref{eqs:q-low-freq} does not violate the conditions $\mathcal P_{\rm TM}(\xi;q)\neq 0$ and $\mathcal P_{\rm TE}(\xi;q)\neq 0$ for all real $\xi$, which are assumed for the application of the underlying Wiener-Hopf factorization in Section~\ref{subsec:EP-disp}. \color{black}

\section{On the nonretarded frequency regime}
\label{sec:qs}
In this section, we simplify~\eqref{eq:EP-dispersion-I} under the conditions
\begin{equation*}
	\biggl|\frac{\omega\mu\sigma(\omega)}{k_0}\biggr|\ll 1\ \mbox{and}\ \Im\,\sigma(\omega)>0~,
\end{equation*}
which signify the nonretarded frequency regime in the context of our isotropic conductivity model.\cite{Pitarkeetal2007,Bludov2013} We show how the quasi-electrostatic approximation of previous works~\cite{VolkovMikhailov1985,VolkovMikhailov1988} can be refined. In fact, we derive a correction to this approximation which indicates the role of the TE polarization through the relatively small $R_\pm(\pm \iu q{\rm sg}(q))$ terms in~\eqref{eq:EP-dispersion-I}. In this regime, we expect to have\cite{VolkovMikhailov1988,Fetter1985}
\begin{equation*}
\eta(q)=-\frac{\iu\omega\mu\sigma}{2k_0}\frac{q}{k_0}=-\frac{\iu\omega\mu\sigma}{2k_0}\frac{1}{\delta}=\mathcal O(1)~,\quad \delta=\frac{k_0}{q}~;
\end{equation*}
thus, $|\delta|\ll 1$ ($|q|\gg |k_0|$). The definition of $\eta(q)$ is inspired by the quasi-electrostatic approach of Ref.~\onlinecite{VolkovMikhailov1988} where it is found that $\eta \simeq 1.217$. We assume that $\Re\,q>0$.

First, we expand in $\delta$ the integral pertaining to $\mathcal P_{\rm TE}$ for fixed $\eta$. By~\eqref{eq:R-int-def} we have
\begin{align*}
R_+(\iu q)+R_-(-\iu q)&=\frac{2}{\pi}\int_0^\infty \frac{{\rm d}\varsigma}{1+\varsigma^2}\ \ln\biggl[1+\eta \delta^2\,(\varsigma^2+1-\delta^2)^{-1/2}\biggr]\\
&\sim \frac{2}{\pi}\int_0^\infty \frac{{\rm d}\varsigma}{1+\varsigma^2}\ \ln\biggl[1+\eta\delta^2\,(1+\varsigma^2)^{-1/2}\biggl(1+\frac{\delta^2}{2}\frac{1}{1+\varsigma^2}\biggr)\biggr]\\
&\sim \frac{2}{\pi}\int_0^\infty \frac{{\rm d}\varsigma}{1+\varsigma^2}\ \biggl\{\eta\delta^2\,(1+\varsigma^2)^{-1/2}\biggl(1+\frac{\delta^2}{2}\frac{1}{1+\varsigma^2}\biggr)-\frac{1}{2}\frac{(\eta\delta^2)^2}{1+\varsigma^2}\biggr\} \\
&=\frac{2}{\pi}\eta\delta^2
\biggl[1+\biggl(\frac{1}{3}- \frac{\pi}{8}\eta\biggr)\delta^2\biggr]~,\quad |\delta|\ll 1~.
\end{align*}

In contrast, the integral pertaining to $\mathcal P_{\rm TM}$ is 
\begin{equation*}
Q_+(\iu q)+Q_-(-\iu q)=\frac{2}{\pi}\int_0^\infty \frac{{\rm d}\varsigma}{1+\varsigma^2}\ \ln\left(1-\eta\sqrt{\varsigma^2+1-\delta^2}\right)=\mathcal O(1)~,
\end{equation*}
which is dominant over $R_+(\iu q)+R_-(-\iu q)$. 
By expanding in $\delta$ for fixed $\eta=\eta(q)$, we find
\begin{align*}
&Q_+(\iu q)+Q_-(-\iu q)\sim \frac{2}{\pi}\int_0^\infty \frac{{\rm d}\varsigma}{1+\varsigma^2}\,\ln\biggl\{1-\eta\sqrt{1+\varsigma^2}\biggl[1-\frac{\delta^2}{2}\frac{1}{1+\varsigma^2}-\frac{\delta^4}{8}\frac{1}{(1+\varsigma^2)^2}\biggr]\biggr\}\\
&\mbox{} \quad \sim \frac{2}{\pi}\biggl\{\int_0^\infty \frac{{\rm d}\varsigma}{1+\varsigma^2}\ \ln\left(1-\eta\sqrt{1+\varsigma^2}\right)-\frac{1}{2}\eta\delta^2\int_0^\infty \frac{{\rm d}\varsigma}{(1+\varsigma^2)^{3/2}}\ \left(\eta\sqrt{1+\varsigma^2}-1\right)^{-1}\\
&\mbox{} \quad -\frac{1}{8}\eta\delta^4\biggl[\eta\int_0^\infty \frac{{\rm d}\varsigma}{(1+\varsigma^2)^2}\,(\eta\sqrt{1+\varsigma^2}-1)^{-2} +\int_0^\infty \frac{{\rm d}\varsigma}{(1+\varsigma^2)^{5/2}}\ \left(\eta\sqrt{1+\varsigma^2}-1\right)^{-1}\biggr]\biggr\}~.
\end{align*}

By neglecting terms $\mathcal O(\delta^4)$, we thus approximate dispersion relation~\eqref{eq:EP-dispersion-I} with $l=0$ by
\begin{equation}\label{eq:EP-disp-appI}
2\int_0^\infty \frac{{\rm d}\varsigma}{1+\varsigma^2}\ln\left(\eta\sqrt{1+\varsigma^2}-1\right)\sim \eta\delta^2\left\{\eta\int_0^\infty \frac{{\rm d}\varsigma}{1+\varsigma^2} \left(\eta\sqrt{1+\varsigma^2}-1\right)^{-1}+1\right\}~,
\end{equation}
where the term $\iu\pi=\ln(-1)$ from~\eqref{eq:EP-dispersion-I} was combined with the logarithm from $Q_+(\iu q)+Q_-(-\iu q)$, resulting in the reversal of the sign of its argument.
Note that $\eta\delta=-\iu\omega\mu\sigma/(2k_0)$. Thus, the solution, $q$, of~\eqref{eq:EP-disp-appI}
is expressed via the expansion
\begin{equation*}
\eta(q)\sim \eta_0 \biggl\{1-\eta_1\biggl(\frac{\omega\mu\sigma}{2k_0}\biggr)^2\biggr\}~,\quad \eta_j=\mathcal O(1)\quad (j=0,\,1)~;\quad \biggl|\frac{\omega\mu\sigma}{2k_0}\biggr|\ll 1~.
\end{equation*}
The coefficients $\eta_j$ are determined below. 
The substitution of the above expansion into~\eqref{eq:EP-disp-appI} along with a dominant balance argument yield the desired equations for $\eta_j$, viz.,
\begin{subequations}\label{eqs:EP-disp-appI}
\begin{equation}\label{eq:EP-disp-qs-zeroth}
\int_0^\infty \frac{{\rm d}\varsigma}{1+\varsigma^2}\ln\left(\eta_0\sqrt{1+\varsigma^2}-1\right)=0~,
\end{equation}
\begin{equation}\label{eq:EP-disp-qs-corr}
2\eta_0\eta_1 \int_0^\infty \frac{{\rm d}\varsigma}{\sqrt{1+\varsigma^2}}\biggl(\eta_0\sqrt{1+\varsigma^2}-1\biggr)^{-1}=\int_0^\infty \frac{{\rm d}\varsigma}{1+\varsigma^2} \biggl(\eta_0\sqrt{1+\varsigma^2}-1\biggr)^{-1}+\eta_0^{-1}~.
\end{equation}
\end{subequations}

Equation~\eqref{eq:EP-disp-qs-zeroth} is in agreement with the  corresponding result in the quasi-electrostatic limit derived in Ref.~\onlinecite{VolkovMikhailov1988} and gives $\eta_0\simeq 1.217$; cf.~their equation (39) and the subsequent relation in view of the change of variable $\varsigma\mapsto x$ with $\varsigma=\cot x$ here. On the other hand, after some algebra \eqref{eq:EP-disp-qs-corr} entails
\begin{equation*}
\eta_1=\frac{1}{2} \left\{1-\sqrt{1-\eta_0^{-2}}\frac{\displaystyle \frac{\pi}{2}-\eta_0^{-1}}{\displaystyle \frac{\pi}{2}+\arcsin(\eta_0^{-1})}\right\}\quad \left(0< \arcsin w<\frac{\pi}{2}\ \mbox{if}\ 0<w<1\right)~.
\end{equation*}
 The numerical evaluation of this coefficient yields $\eta_1\simeq 0.416$ by use of $\eta_0\simeq 1.217$. According to our expansion for $\eta(q)$ above, the EP wave number is furnished by  
 \begin{equation}\label{eq:q-qs-asympt}
 q= \eta_0\,\frac{\iu 2k_0^2}{\omega\mu\sigma}\biggl\{1-\eta_1\biggl(\frac{\omega\mu\sigma}{2k_0}\biggr)^2+
 \mathcal O\biggl(\biggl(\frac{\omega\mu\sigma}{2k_0}\biggr)^4\biggr)\biggr\}~.
 \end{equation}
Evidently, the leading-order correction term, which is of the relative order of $(\omega\mu\sigma/k_0)^2$, comes from contributions of both TM and TE polarizations, i.e., from both the $Q_\pm$ and $R_\pm$ terms in dispersion relation~\eqref{eq:EP-dispersion-I}. 
 
 By virtue of~\eqref{eq:q-qs-asympt}, it is of some interest to describe $q(\omega)$ when $\sigma(\omega)$ is given by the Drude model which is relevant to doped single-layer graphene for small enough plasmonic energies (see Section~\ref{sec:small-q}).\cite{Bludov2013,Jablan2013} By use of the formula $\sigma(\omega)=\iu (\mathcal D/\pi)\,(\omega+\iu/\tau_e)^{-1}$ (beginning of Section~\ref{sec:small-q}), where the dimensional parameter $\mathcal D$ is the Drude weight,\cite{Jablan2013} we obtain
\begin{align*}
\Re\,q(\omega)&\sim \frac{2\eta_0\varepsilon\pi}{\mathcal D}\omega^2 \biggl\{1+\frac{\eta_1}{4\pi^2}\frac{(Z_0 \mathcal D)^2}{\omega^2+\tau_e^{-2}}\biggr\}\sim \frac{2\eta_0\varepsilon\pi}{\mathcal D} \biggl\{\omega^2+\frac{\eta_1}{4\pi^2}(Z_0 \mathcal D)^2\biggr\} ~,\\
\Im\,q(\omega)&\sim \frac{2\eta_0\varepsilon\pi}{\mathcal D} \omega\tau_e^{-1}\biggl\{1-\frac{\eta_1}{4\pi^2}\frac{(Z_0 \mathcal D)^2}{\omega^2+\tau_e^{-2}}\biggr\}\sim \frac{2\eta_0\varepsilon\pi}{\mathcal D} \omega\tau_e^{-1}\biggl\{1-\frac{\eta_1}{4\pi^2}\frac{(Z_0 \mathcal D)^2}{\omega^2}\biggr\} ~.
\end{align*}
Here, the formulas on the rightmost-hand side come from applying the condition $\omega\tau_e\gg 1$.

\section{Extension: Two coplanar conducting sheets}
\label{sec:extension}
In this section, we extend our formalism to the setting with two coplanar, semi-infinite sheets of distinct isotropic and homogeneous conductivities. Consider the `left' sheet $\Sigma^L=\{(x,y,z)\in\mathbb{R}^3\,:\,z=0,\, x<0\}$ and the `right' sheet $\Sigma^R=\{(x,y,z)\in\mathbb{R}^3\,:\,z=0,\, x>0\}$ that have scalar, spatially constant surface conductivities $\sigma^L(\omega)$ and $\sigma^R(\omega)$, respectively ($\sigma^{L}\neq \sigma^{R}$ and $\sigma^L\sigma^R\neq 0$). We formulate and solve a system of Wiener-Hopf integral equations for the electric field tangential to the plane of the sheets on $\Sigma=\Sigma^L\cup \Sigma^R$ \color{black} in order to derive the dispersion relation for the EP that propagates along the $y$-axis.

The surface current density is $\bjs(x,y)=e^{\iu q y}\sigma(x) \{E_x(x,z) \ex +E_y(x,z)\ey\}\bigl|_{z=0}$ where 
\begin{equation*}
	\sigma(x)=\sigma^L+ \vartheta(x) (\sigma^R-\sigma^L)\qquad (\sigma^R\neq \sigma^L)~;
\end{equation*}
the Heaviside step function $\vartheta(x)$ is defined by $\vartheta(x)=1$ if $x>0$ and $\vartheta(x)=0$ if $x<0$. 
By using the vector potential in the Lorenz gauge (Section~\ref{subsec:integral}),\cite{King1963} we obtain the system
\begin{equation*}
\begin{pmatrix} u(x) \\
                v(x) 
\end{pmatrix} =    \frac{\iu\omega\mu}{k_0^2}
\begin{pmatrix}
\displaystyle \frac{{\rm d}^2}{\dx^2}+k_0^2 &  \quad \displaystyle \iu q \frac{{\rm d}}{\dx} \\
  \displaystyle \iu q\frac{{\rm d}}{\dx}      &   k_{\rm eff}^2   	
\end{pmatrix} 
\int_{-\infty}^\infty \dxp\, K(x-x')\sigma(x')\,  
\begin{pmatrix} u(x') \\
                v(x') 
\end{pmatrix} \quad  x\ \mbox{in}\ \mathbb{R}~,            
\end{equation*}
where $u(x)=E_x(x,0)$ and $v(x)=E_y(x,0)$.
By applying the Fourier transform with respect to $x$, we obtain the functional equations [cf.~\eqref{eq:integral-rels-FT-const}]
\begin{subequations}\label{eqs:func-2sheets}
\begin{equation}\label{eq:RH-2sheets}
\bkm^R(\xi)
\begin{pmatrix} \widehat u_{-}(\xi) \\
                \widehat v_{-}(\xi) 
\end{pmatrix}
+\bkm^L(\xi)
\begin{pmatrix} \widehat u_{+}(\xi) \\
                \widehat v_{+}(\xi) 
\end{pmatrix}  
=0\qquad (\mbox{all\ real}\ \xi)~,          	
\end{equation}
where
\begin{equation}\label{eq:Lambda-rightleft-2sheets}
\bkm^{\ell}(\xi):= \begin{pmatrix}
\displaystyle 1-\frac{\iu\omega\mu\sigma^{\ell}}{k_0^2}(k_0^2-\xi^2)\widehat K(\xi;q) &  \ \displaystyle -\frac{\iu\omega\mu\sigma^{\ell}}{k_0^2}(\iu q)(\iu \xi) \widehat K(\xi;q)  \\
  \displaystyle -\frac{\iu\omega\mu\sigma^{\ell}}{k_0^2}(\iu q)(\iu\xi)\widehat K(\xi; q)      &   \ \displaystyle 1-\frac{\iu\omega\mu\sigma^{\ell}}{k_0^2}(k_0^2-q^2) \widehat K(\xi;q)   	\end{pmatrix}~;\quad \ell=R,\,L~.  
\end{equation}
\end{subequations}

In the spirit of our analysis for a single sheet (Section~\ref{subsec:explicit}), we diagonalize the matrices $\bkm^{\ell}(\xi)$. Their eigenvalues are [cf.~\eqref{eq:K-eigenv}]
\begin{equation*}
\mathcal P_{\rm TM}^{\ell}(\xi)=1-\frac{\iu\omega\mu\sigma^{\ell}}{k_0^2}(k_{\rm eff}^2-\xi^2)\widehat K(\xi)~,\quad \mathcal P_{\rm TE}^{\ell}(\xi)=1-\iu\omega\mu\sigma^{\ell}\widehat K(\xi)\qquad (\ell=R,\,L)~.
\end{equation*}
Recall that $\widehat K(\xi)=(\iu/2)(k_{\rm eff}^2-\xi^2)^{-1/2}$ where $k_{\rm eff}=\sqrt{k_0^2-q^2}$; $\Im\sqrt{k_{\rm eff}^2-\xi^2}>0$ for the first Riemann sheet.
Accordingly, functional equations~\eqref{eq:RH-2sheets} become
\begin{equation*}
\begin{pmatrix}
\mathcal P_{\rm TM}^L(\xi) & 0 \\
0  & \mathcal P_{\rm TE}^L(\xi) 
\end{pmatrix}
\boldsymbol{\mathcal S}(\xi)
\begin{pmatrix}
\widehat u_+(\xi) \\
\widehat v_+(\xi) 
\end{pmatrix}
+
\begin{pmatrix}
\mathcal P_{\rm TM}^R(\xi) & 0 \\
0  & \mathcal P_{\rm TE}^R(\xi) 
\end{pmatrix}
\boldsymbol{\mathcal S}(\xi)
\begin{pmatrix}
\widehat u_-(\xi) \\
\widehat v_-(\xi) 
\end{pmatrix}=0~,
\end{equation*}
for real $\xi$, where the matrix $\boldsymbol{\mathcal S}(\xi)$ is defined by~\eqref{eq:S-matrix-def}.
Hence, we recover and apply transformation~\eqref{eq:UV-transf-def} where $(\widehat u, \widehat v)\mapsto (U,V)$; the functions $U_\pm(\xi)$ and $V_\pm(\xi)$ satisfy the equations
\begin{align*}
& \mathcal P_{\rm TM}^L(\xi)U_+(\xi)+\mathcal P_{\rm TM}^R(\xi)U_-(\xi)=0~,\\
&\mathcal P_{\rm TE}^L(\xi)V_+(\xi)+\mathcal P_{\rm TE}^R(\xi)V_-(\xi)=0\qquad \mbox{all\ real}\ \xi~. 
\end{align*}
These equations form an extension of~\eqref{eq:transf-func-eqs} to the geometry of two coplanar sheets. 

Following the procedure of Section~\ref{subsec:EP-disp}, we assume that
\begin{equation*}
\mathcal P_{\rm TM}^{\ell}(\xi)\neq 0\ \mbox{and}\ \mathcal P_{\rm TE}^{\ell}(\xi)\neq 0\quad \mbox{for\ all\ real}\ \xi\qquad (\ell=R,\,L)~.
\end{equation*}
By defining the functions
\begin{equation}\label{eq:P-TME-2sheets-def}
\mathcal P_{\rm TM}(\xi)= \frac{\mathcal P_{\rm TM}^R(\xi)}{\mathcal P_{\rm TM}^L(\xi)}~,\qquad \mathcal P_{\rm TE}(\xi)= \frac{\mathcal P_{\rm TE}^R(\xi)}{\mathcal P_{\rm TE}^L(\xi)}
\end{equation}
with $Q(\xi)=\ln\mathcal P_{\rm TM}(\xi)$ and $R(\xi)=\ln\mathcal P_{\rm TE}(\xi)$, we carry out the splittings indicated in~\eqref{eq:QR-split-def}; the logarithmic functions here are such that $Q(\xi)=\ln\mathcal P_{\rm TM}^R(\xi)-\ln\mathcal P_{\rm TM}^L(\xi)$ and $R(\xi)=\ln\mathcal P_{\rm TE}^R(\xi)-\ln\mathcal P_{\rm TE}^L(\xi)$ when $\xi$ lies in the top Riemann sheet (cf. Remark~3). Because in the present setting the indices associated with $\mathcal P_{\rm TM}(\xi)$ and $\mathcal P_{\rm TE}(\xi)$ on the real axis are zero, i.e., $\boldsymbol\nu=0$ as in~\eqref{eq:index-zero}, the split functions $Q_\pm(\xi)$ and $R_\pm(\xi)$ are given by integrals~\eqref{eq:QR-def} under~\eqref{eq:P-TME-2sheets-def}. Note that $e^{Q_\pm(\xi)}=\mathcal O(1)$ and $e^{R_\pm(\xi)}\to 1$ as $\xi\to\infty$ in $\mathbb{C}_\pm$; see~Appendix. The Wiener-Hopf method furnishes the entire functions $\mathcal E_1(\xi)=C_1={\rm const.}$ and $\mathcal E_2(\xi)=C_2={\rm const.}$, as in the case with a single conducting sheet (Section~\ref{subsec:EP-disp}). Some intermediate steps are slightly different because of the asymptotics for $e^{Q_\pm(\xi)}$ in the setting with two sheets (see Appendix). We omit any further details about how to obtain $\mathcal E_1(\xi)$ and $\mathcal E_2(\xi)$ here.

Consequently, we obtain the formulas
$U_\pm(\xi)=\mp C_1 e^{\pm Q_\pm(\xi)}$ and $V_\pm(\xi)=\mp C_2 e^{\pm R_+(\xi)}$ where $C_1$ and $C_2$ are arbitrary constants, which in turn yield~\eqref{eq:uv-soln-minus} and~\eqref{eq:uv-soln-plus} for $\widehat u_\pm(\xi)$ and $\widehat v_\pm(\xi)$.
By the analyticity of $u_-(\xi)$ and $v_-(\xi)$ at $\xi=-\iu q{\rm sg}(q)$, and the analyticity of $u_+(\xi)$ and $v_+(\xi)$ at $\xi=\iu q{\rm sg}(q)$, we subsequently derive relations~\eqref{eqs:dispersion}. 


\section{Conclusion and discussion}
\label{sec:conclusion}
In this paper, by using the theory of the Wiener-Hopf integral equations we derived the dispersion relation for the edge plasmon-polariton that propagates along the straight edge of a semi-infinite, planar conducting sheet. The sheet lies in a uniform isotropic medium. Our treatment takes into account retardation effects, in the sense that, given a spatially homogeneous scalar conductivity of the 2D material as a function of frequency, the underlying boundary value of Maxwell's equations is solved exactly. Thus, we avoid the restrictive assumptions of the quasi-electrostatic approximation. Our formalism was directly extended to the geometry with two semi-infinite, coplanar conducting sheets. 

In our formal analysis, the existence of the EP dispersion relation on the isotropic sheet is connected to the notion of {\em zero index} in Krein's theory.\cite{Krein1962} In the setting of the dissipationless Drude model for the surface conductivity,\cite{Jablan2013} for example, this zero index mathematically expresses the property that, for every (real) EP wave number $q$, the corresponding EP frequency, or energy, $\omega(q)$ is smaller than the energy of the 2D bulk SP of the same wave number. Thus, the character of this EP remains intact in the isotropic setting, in contrast to the situation with a strictly anisotropic conductivity, e.g., in the presence of a static magnetic field, where a branch of $\omega(q)$ may cross the respective dispersion curve of the 2D bulk SP.\cite{VolkovMikhailov1988} This latter possibility is studied in some generality, yet within the quasi-electrostatic approach, elsewhere.\cite{MMSLL-preprint,MMSLL-inprep}

The EP dispersion relation derived here expresses the simultaneous presence of distinct polarization effects. To be more precise, the effect of the TM polarization, which alone provides the fine scale of the bulk SP in the nonretarded frequency regime, is accompanied by a contribution that amounts to the TE polarization. In this framework, we were able to smoothly connect two non-overlapping asymptotic regimes: (i) the low-frequency limit, in which the EP wave number, $q$, approaches the free-space propagation constant, $k_0$, and thus $q/\omega \sim {\rm const.}$; and (ii) the nonretarded frequency regime, where $q$ is much larger in magnitude than $k_0$ and $q/\omega^2 \sim {\rm const.}$ In each of these regimes, we derived corrections to the anticipated, leading-order formulas for $q(\omega)$ by invoking the semi-classical Drude model.

Our work has limitations and leaves several open questions. Two noteworthy issues are the stability of the EP under perturbations of the edge \color{black} and the semi-infinite character of the sheet geometry. As a next step, it is tempting to analyze the EP dispersion in microstrips, which may be more closely related to the actual experimental setups.\cite{Feietal2015,Taoetal2011} This setting calls for developing approximate solution schemes for the related integral equations for the electric field. Since we addressed only isotropic and homogeneous surface conductivities, it is natural to investigate how to analyze anisotropic or nonhomogeneous sheets with nonlocalities.\cite{MMSLL-preprint} In this context, a possibility is to couple the full Maxwell equations with linearized models of viscous electron flow in the hydrodynamic regime,\cite{Lucas2018} where the viscosity and compressibility induce nonlocal effects in the effective conductivity tensor within linear response theory; moreover, the edge as a boundary of the viscous 2D electron system necessarily affects the form of the conductivity tensor. 

\section*{ACKNOWLEDGMENTS}

The author is indebted to Vera Andreeva, Tony Low, Alex Levchenko, Andy Lucas, Mitchell Luskin, Matthias Maier, Marco Polini, Tobias Stauber, and Tai Tsun Wu for useful discussions. The author also acknowledges: partial support by the MURI Award No.~W911NF-14-1-0247 of the Army Research Office (ARO) and Grant No.~1517162 of the Division of Mathematical Sciences (DMS) of the NSF; the support by a Research and Scholarship Award from the Graduate School, University of Maryland in the spring of 2019; and the support of the Institute for Mathematics and its Applications (NSF Grant DMS-1440471) at the University of Minnesota for several visits.

\appendix*

\section{On asymptotic expansions for $Q_\pm(\xi)$ and $R_\pm(\xi)$ as $\xi\to\infty$}
\label{app:QR-asympt}
In this appendix, we sketch the derivations of asymptotic formulas for the split functions $Q_\pm(\xi)$ and $R_\pm(\xi)$ as $\xi\to\infty$ in $\mathbb{C}_\pm$ (see Sections~\ref{sec:solution} and~\ref{sec:extension}). For analogous  asymptotic expansions, see Refs. \onlinecite{MML2017,MMSLL-preprint}.
\medskip

\subsection{Single conducting sheet} 
\label{subsec:asympt-1sheet}

Consider formulas~\eqref{eq:QR-def} for $Q_\pm(\xi)$ and $R_\pm(\xi)$ with the functions $\mathcal P_{\rm TM}(\xi)$ and $\mathcal P_{\rm TE}(\xi)$ introduced in~\eqref{eq:K-eigenv}. We express the associated integrals in the forms
\begin{equation*}
Q_\pm(\xi)=\pm\frac{1}{\iu\pi}\int_0^{\infty e^{-\iu\arg\xi}} {\rm d}\varsigma\ \frac{Q(\xi\varsigma)}{\varsigma^2-1}~,\quad 
R_\pm(\xi)=\pm \frac{1}{\iu\pi}\int_0^{\infty e^{-\iu\arg\xi}}
{\rm d}\varsigma\ \frac{R(\xi\varsigma)}{\varsigma^2-1}\quad (\pm\Im\,\xi>0)~,
\end{equation*}
where
\begin{equation*}
Q(\zeta)=\ln\biggl(1+\frac{\iu\omega\mu\sigma}{2k_0^2}\sqrt{\zeta^2-k_{\rm eff}^2}\biggr)~,\quad 
R(\zeta)=\ln\biggl(1-\frac{\iu\omega\mu\sigma}{2}\frac{1}{\sqrt{\zeta^2-k_{\rm eff}^2}}\biggr)~;\quad \Re\sqrt{\zeta^2-k_{\rm eff}^2}>0~.
\end{equation*}
First, let us focus on $Q_+(\xi)$. The numerator in the corresponding integrand is expressed as
\begin{equation*}
Q(\xi\varsigma)=\ln\biggl(\frac{\iu\omega\mu\sigma}{2k_0^2}\xi\varsigma\biggr)+Q_1(\xi\varsigma)~;\quad 
Q_1(\zeta)=\ln\biggl(\sqrt{1-\frac{k_{\rm eff}^2}{\zeta^2}}+\frac{2k_0^2}{\iu\omega\mu\sigma}\frac{1}{\zeta}\biggr)~.
\end{equation*}
Notice that $Q_1(\zeta)=\mathcal O(\zeta^{-1})$ as $\zeta\to\infty$. Thus, by substitution of this $Q(\xi\varsigma)$ into the integral for $Q_+(\xi)$ and exact evaluation of the contribution of the first term, we obtain  \cite{MML2017,MMSLL-preprint}
\begin{equation}\label{eq:Q+_asympt-xi}
Q_+(\xi)=\frac{1}{2}\ln\biggl(\frac{\omega\mu\sigma\xi}{2k_0^2}\biggr)+\mathcal O\biggl(\frac{1+\ln\xi}{\xi}\biggr)\quad \mbox{as}\ \xi\to\infty\ \mbox{in}\ \mathbb{C}_+~;
\end{equation}
the correction term can be systematically derived via the Mellin transform technique.\cite{Sasiela1993}
In the above asymptotic formula for $Q_+(\xi)$, the branch cut for the logarithm can lie in the lower half $\xi$-plane or the negative real axis.
By symmetry, we have
\begin{equation}\label{eq:Q-_asympt-xi}
Q_-(\xi)=\frac{1}{2}\ln\biggl(-\frac{\omega\mu\sigma\xi}{2k_0^2}\biggr)+\mathcal O\biggl(\frac{1+\ln\xi}{\xi}\biggr)\quad \mbox{as}\ \xi\to\infty\ \mbox{in}\ \mathbb{C}_-~,
\end{equation}
where the branch cut for the logarithm can lie in the upper half $\xi$-plane or the negative real axis. To reconcile the last  two asymptotic formulas for $Q_+(\xi)$ and $Q_-(\xi)$, we take the branch cut for each logarithm along the negative real axis. Accordingly, we verify that
\begin{equation*}
Q_+(\xi)+Q_-(\xi)\sim \ln\biggl(\frac{\iu\omega\mu\sigma\xi}{2k_0^2}\biggr)\sim Q(\xi)\quad \mbox{as}\ \xi\to\infty\quad \mbox{in}\ \mathbb{C}_+\ \mbox{and}\ \mathbb{C}_-~.
\end{equation*}
We now turn our attention to $R_+(\xi)$. We write 
\begin{equation}\label{eq:R-xi-1sheet}
R(\xi\varsigma)=\ln\biggl(1-\frac{\iu\omega\mu\sigma}{2\xi\varsigma}\biggr)+R_1(\xi\varsigma)~;\quad R_1(\zeta)=\ln\biggl\{1-\frac{\iu\omega\mu\sigma}{2\zeta}\frac{(1-k_{\rm eff}^2/\zeta^2)^{-1/2}-1}{1-\iu\omega\mu\sigma/(2\zeta)}\biggr\}~,
\end{equation}
where $R_1(\zeta)=\mathcal O(\zeta^{-3})$ as $\zeta\to \infty$. The substitution of the above expression for $R(\xi\varsigma)$ into the integral for $R_+(\xi)$ yields
\begin{equation}\label{eq:R+_asympt-xi}
R_+(\xi)=\frac{1}{\pi} \frac{\omega\mu\sigma}{2\xi}\,\ln\biggl(\frac{2\xi}{\omega\mu\sigma}\biggr)+\mathcal O(1/\xi)\quad \mbox{as}\ \xi\to\infty\ \mbox{in}\ \mathbb{C}_+~.
\end{equation}
In the last formula, the logarithm comes from the first term shown in~\eqref{eq:R-xi-1sheet}; while the $\mathcal O(1/\xi)$ correction term is attributed to both the first and second terms appearing in~\eqref{eq:R-xi-1sheet}.
Similarly, we have 
\begin{equation}\label{eq:R-_asympt-xi}
R_-(\xi)=-\frac{1}{\pi} \frac{\omega\mu\sigma}{2\xi}\,\ln\biggl(-\frac{2\xi}{\omega\mu\sigma}\biggr)+\mathcal O(1/\xi)\quad \mbox{as}\ \xi\to\infty\ \mbox{in}\ \mathbb{C}_-~.
\end{equation}
We note in passing that $R_+(\xi)+R_-(\xi)=\mathcal O(1/\xi)$ as $\xi\to\infty$, as expected because the sum of $R_+(\xi)$ and $R_-(\xi)$ should be exactly equal to $R(\xi)$. 
\medskip

\subsection{Two coplanar conducting sheets} 
\label{subsec:asympt-2sheets}

Consider integral formulas~\eqref{eq:QR-def} for $Q_\pm(\xi)$ and $R_\pm(\xi)$ where the functions $\mathcal P_{\rm TM}(\xi)$ and $\mathcal P_{\rm TE}(\xi)$ are now defined by~\eqref{eq:P-TME-2sheets-def} (Section~\ref{sec:extension}). The EP is assumed to propagate along the joint boundary of two coplanar sheets of distinct, scalar surface conductivities $\sigma^R$ and $\sigma^L$ with $\sigma^R\neq \sigma^L$ and $\sigma^R\sigma^L\neq 0$. For this geometry, we have
\begin{equation*}
Q(\zeta)=\ln\frac{\mathcal P_{\rm TM}^R(\zeta)}{\mathcal P_{\rm TM}^L(\zeta)}=\ln \mathcal P_{\rm TM}^R(\zeta)-\ln\mathcal P_{\rm TM}^L(\zeta)=\ln\biggl(\frac{\sigma^R}{\sigma^L}\biggr)+\mathcal O(1/\zeta)\quad \mbox{as}\ \zeta\to\infty
\end{equation*}
and
\begin{equation*}
R(\zeta)=\ln\frac{\mathcal P_{\rm TE}^R(\zeta)}{\mathcal P_{\rm TE}^L(\zeta)} =\ln\mathcal P_{\rm TE}^R(\zeta)-\ln\mathcal P_{\rm TE}^L(\zeta)=\mathcal O(1/\zeta)\quad \mbox{as}\ \zeta\to \infty~,
\end{equation*}
in the appropriately chosen branch of the logarithm, $w=\ln\mathcal P_\varpi^\ell$ ($\varpi={\rm TM}, {\rm TE}$ and $\ell=R, L$).
By inspection of the resulting integrals for $Q_\pm(\xi)$ and $R_\pm(\xi)$ here we realize that their treatment for a single sheet in Section~\ref{subsec:asympt-1sheet} of this Appendix can be directly applied to the present setting of two sheets. Without further ado, in regard to $R_\pm(\xi)$ we can assert that
\begin{equation}\label{eq:R+-_asympt-xi-2sheets}
R_\pm(\xi)=\pm\frac{1}{\pi} \frac{\omega\mu}{2\xi}\biggl\{\sigma^R\ln\biggl(\pm\frac{2\xi}{\omega\mu\sigma^R}\biggr)-\sigma^L\ln\biggl(\pm\frac{2\xi}{\omega\mu\sigma^L}\biggr)\biggr\} +\mathcal O(1/\xi)\quad \mbox{as}\ \xi\to\infty\ \mbox{in}\ \mathbb{C}_\pm~;
\end{equation}
thus, $R_\pm(\xi)=o(1)$. On the other hand, in regard to the asymptotics for $Q_\pm(\xi)$ we find
\begin{equation}\label{eq:Q+-_asympt-xi-2sheets}
Q_\pm(\xi)=\frac{1}{2}\ln\biggl(\frac{\sigma^R}{\sigma^L}\biggr)+\mathcal O\biggl(\frac{1+\ln\,\xi}{\xi}\biggr)\quad \mbox{as}\ \xi\to\infty\ \mbox{in}\ \mathbb{C}_\pm~,
\end{equation}
with $\sigma^R\neq \sigma^L$ and $\sigma^L\sigma^R\neq 0$.


\end{document}